\magnification=\magstep1 
\font\bigbfont=cmbx10 scaled\magstep1
\font\bigifont=cmti10 scaled\magstep1
\font\bigrfont=cmr10 scaled\magstep1
\vsize = 23.5 truecm
\hsize = 15.5 truecm
\hoffset = .2truein
\baselineskip = 14 truept
\overfullrule = 0pt
\parskip = 3 truept
\def\one{{\mathchoice {1\mskip-4mu {\text {\rm l}}} {
1\mskip-4mu {\text {\rm l}}} {1\mskip-4.5mu {\text {\rm l}}} {1\mskip-5mu
{\text {\rm l}}}}}
\def\fmtname{AmS-TeX}

\def\fmtversion{2.1}
\catcode`\@=11
\ifx\amstexloaded@\relax\catcode`\@=\active
   \else\let\amstexloaded@\relax\fi
\newlinechar=`\^^J
\def\W@{\immediate\write\sixt@@n}
\def\CR@{\W@{^^J\fmtname - Version \fmtversion^^J%
COPYRIGHT 1985, 1990, 1991 - AMERICAN MATHEMATICAL SOCIETY^^J%
Use of this macro package is not restricted provided^^J%
each use is acknowledged upon publication.^^J}}
\CR@ \everyjob{\CR@}
\message{Loading definitions for}
\message{misc utility macros,}
\toksdef\toks@@=2
\long\def\rightappend@#1\to#2{\toks@{\\{#1}}\toks@@
 =\expandafter{#2}\xdef#2{\the\toks@@\the\toks@}\toks@{}\toks@@{}}
\def\alloclist@{}
\newif\ifalloc@
\def\showallocations{{\def\\{\immediate\write\m@ne}\alloclist@}\alloc@true}
\def\alloc@#1#2#3#4#5{\global\advance\count1#1by\@ne
 \ch@ck#1#4#2\allocationnumber=\count1#1
 \global#3#5=\allocationnumber
 \edef\next@{\string#5=\string#2\the\allocationnumber}%
 \expandafter\rightappend@\next@\to\alloclist@}
\newcount\count@@
\newcount\count@@@
\def\FN@{\futurelet\next}
\def\DN@{\def\next@}
\def\DNii@{\def\nextii@}
\def\RIfM@{\relax\ifmmode}
\def\RIfMIfI@{\relax\ifmmode\ifinner}
\def\setboxz@h{\setbox\z@\hbox}
\def\wdz@{\wd\z@}
\def\boxz@{\box\z@}
\def\setbox@ne{\setbox\@ne}
\def\wd@ne{\wd\@ne}
\def\iterate{\body\expandafter\iterate\else\fi}
\def\err@#1{\errmessage{AmS-TeX error: #1}}
\newhelp\defaulthelp@{Sorry, I already gave what help I could...^^J
Maybe you should try asking a human?^^J
An error might have occurred before I noticed any problems.^^J
``If all else fails, read the instructions.''}
\def\Err@{\errhelp\defaulthelp@\err@}
\def\eat@#1{}
\def\in@#1#2{\def\in@@##1#1##2##3\in@@{\ifx\in@##2\in@false\else\in@true\fi}%
 \in@@#2#1\in@\in@@}
\newif\ifin@
\def\space@.{\futurelet\space@\relax}
\space@. %
\newhelp\athelp@
{Only certain combinations beginning with @ make sense to me.^^J
Perhaps you wanted \string\@\space for a printed @?^^J
I've ignored the character or group after @.}
{\catcode`\~=\active 
 \lccode`\~=`\@ \lowercase{\gdef~{\FN@\at@}}}
\def\at@{\let\next@\at@@
 \ifcat\noexpand\next a\else\ifcat\noexpand\next0\else
 \ifcat\noexpand\next\relax\else
   \let\next\at@@@\fi\fi\fi
 \next@}
\def\at@@#1{\expandafter
 \ifx\csname\space @\string#1\endcsname\relax
  \expandafter\at@@@ \else
  \csname\space @\string#1\expandafter\endcsname\fi}
\def\at@@@#1{\errhelp\athelp@ \err@{\Invalid@@ @}}
\def\atdef@#1{\expandafter\def\csname\space @\string#1\endcsname}
\newhelp\defahelp@{If you typed \string\define\space cs instead of
\string\define\string\cs\space^^J
I've substituted an inaccessible control sequence so that your^^J
definition will be completed without mixing me up too badly.^^J
If you typed \string\define{\string\cs} the inaccessible control sequence^^J
was defined to be \string\cs, and the rest of your^^J
definition appears as input.}
\newhelp\defbhelp@{I've ignored your definition, because it might^^J
conflict with other uses that are important to me.}
\def\define{\FN@\define@}
\def\define@{\ifcat\noexpand\next\relax
 \expandafter\define@@\else\errhelp\defahelp@                               
 \err@{\string\define\space must be followed by a control
 sequence}\expandafter\def\expandafter\nextii@\fi}                          
\def\undefined@@@@@@@@@@{}
\def\preloaded@@@@@@@@@@{}
\def\next@@@@@@@@@@{}
\def\define@@#1{\ifx#1\relax\errhelp\defbhelp@                              
 \err@{\string#1\space is already defined}\DN@{\DNii@}\else
 \expandafter\ifx\csname\expandafter\eat@\string                            
 #1@@@@@@@@@@\endcsname\undefined@@@@@@@@@@\errhelp\defbhelp@
 \err@{\string#1\space can't be defined}\DN@{\DNii@}\else
 \expandafter\ifx\csname\expandafter\eat@\string#1\endcsname\relax          
 \global\let#1\undefined\DN@{\def#1}\else\errhelp\defbhelp@
 \err@{\string#1\space is already defined}\DN@{\DNii@}\fi
 \fi\fi\next@}

\def\predefine#1#2{\let#1#2}
\def\undefine#1{\let#1\undefined}
\message{page layout,}
\newdimen\captionwidth@
\captionwidth@\hsize
\advance\captionwidth@-1.5in
\def\pagewidth#1{\hsize#1\relax
 \captionwidth@\hsize\advance\captionwidth@-1.5in}

\def\hcorrection#1{\advance\hoffset#1\relax}
\def\vcorrection#1{\advance\voffset#1\relax}
\message{accents/punctuation,}

\let\graveaccent\`
\let\acuteaccent\'
\let\tildeaccent\~
\let\hataccent\^
\let\underscore\_
\let\B\=
\let\D\.
\let\ic@\/
\def\/{\unskip\ic@}
\def\textfonti{\the\textfont\@ne}
\def\t#1#2{{\edef\next@{\the\font}\textfonti\accent"7F \next@#1#2}}
\def~{\unskip\nobreak\ \ignorespaces}
\def\.{.\spacefactor\@m}
\atdef@;{\leavevmode\null;}
\atdef@:{\leavevmode\null:}
\atdef@?{\leavevmode\null?}
\edef\@{\string @}
\def\relaxnext@{\let\next\relax}
\atdef@-{\relaxnext@\leavevmode
 \DN@{\ifx\next-\DN@-{\FN@\nextii@}\else
  \DN@{\leavevmode\hbox{-}}\fi\next@}%
 \DNii@{\ifx\next-\DN@-{\leavevmode\hbox{---}}\else
  \DN@{\leavevmode\hbox{--}}\fi\next@}%
 \FN@\next@}
\def\srdr@{\kern.16667em}
\def\drsr@{\kern.02778em}
\def\sldl@{\drsr@}
\def\dlsl@{\srdr@}
\atdef@"{\unskip\relaxnext@
 \DN@{\ifx\next\space@\DN@. {\FN@\nextii@}\else
  \DN@.{\FN@\nextii@}\fi\next@.}%
 \DNii@{\ifx\next`\DN@`{\FN@\nextiii@}\else
  \ifx\next\lq\DN@\lq{\FN@\nextiii@}\else
  \DN@####1{\FN@\nextiv@}\fi\fi\next@}%
 \def\nextiii@{\ifx\next`\DN@`{\sldl@``}\else\ifx\next\lq
  \DN@\lq{\sldl@``}\else\DN@{\dlsl@`}\fi\fi\next@}%
 \def\nextiv@{\ifx\next'\DN@'{\srdr@''}\else
  \ifx\next\rq\DN@\rq{\srdr@''}\else\DN@{\drsr@'}\fi\fi\next@}%
 \FN@\next@}

\def\textfontii{\the\textfont\tw@}
\def\lbrace@{\delimiter"4266308 }
\def\rbrace@{\delimiter"5267309 }
\def\{{\RIfM@\lbrace@\else{\textfontii f}\spacefactor\@m\fi}
\def\}{\RIfM@\rbrace@\else
 \let\@sf\empty\ifhmode\edef\@sf{\spacefactor\the\spacefactor}\fi
 {\textfontii g}\@sf\relax\fi}
\let\lbrace\{
\let\rbrace\}
\def\AmSTeX{{\textfontii A\kern-.1667em%
  \lower.5ex\hbox{M}\kern-.125emS}-\TeX}
\message{line and page breaks,}
\def\vmodeerr@#1{\Err@{\string#1\space not allowed between paragraphs}}
\def\mathmodeerr@#1{\Err@{\string#1\space not allowed in math mode}}
\def\linebreak{\RIfM@\mathmodeerr@\linebreak\else
 \ifhmode\unskip\unkern\break\else\vmodeerr@\linebreak\fi\fi}

\newskip\saveskip@
\def\allowlinebreak{\RIfM@\mathmodeerr@\allowlinebreak\else
 \ifhmode\saveskip@\lastskip\unskip
 \allowbreak\ifdim\saveskip@>\z@\hskip\saveskip@\fi
 \else\vmodeerr@\allowlinebreak\fi\fi}
\def\nolinebreak{\RIfM@\mathmodeerr@\nolinebreak\else
 \ifhmode\saveskip@\lastskip\unskip
 \nobreak\ifdim\saveskip@>\z@\hskip\saveskip@\fi
 \else\vmodeerr@\nolinebreak\fi\fi}
\def\newline{\relaxnext@
 \DN@{\RIfM@\expandafter\mathmodeerr@\expandafter\newline\else
  \ifhmode\ifx\next\par\else
  \expandafter\unskip\expandafter\null\expandafter\hfill\expandafter\break\fi
  \else
  \expandafter\vmodeerr@\expandafter\newline\fi\fi}%
 \FN@\next@}
\def\dmatherr@#1{\Err@{\string#1\space not allowed in display math mode}}
\def\nondmatherr@#1{\Err@{\string#1\space not allowed in non-display math
 mode}}
\def\onlydmatherr@#1{\Err@{\string#1\space allowed only in display math mode}}
\def\nonmatherr@#1{\Err@{\string#1\space allowed only in math mode}}
\def\mathbreak{\RIfMIfI@\break\else
 \dmatherr@\mathbreak\fi\else\nonmatherr@\mathbreak\fi}
\def\nomathbreak{\RIfMIfI@\nobreak\else
 \dmatherr@\nomathbreak\fi\else\nonmatherr@\nomathbreak\fi}
\def\allowmathbreak{\RIfMIfI@\allowbreak\else
 \dmatherr@\allowmathbreak\fi\else\nonmatherr@\allowmathbreak\fi}
\def\pagebreak{\RIfM@
 \ifinner\nondmatherr@\pagebreak\else\postdisplaypenalty-\@M\fi
 \else\ifvmode\removelastskip\break\else\vadjust{\break}\fi\fi}
\def\nopagebreak{\RIfM@
 \ifinner\nondmatherr@\nopagebreak\else\postdisplaypenalty\@M\fi
 \else\ifvmode\nobreak\else\vadjust{\nobreak}\fi\fi}
\def\nonvmodeerr@#1{\Err@{\string#1\space not allowed within a paragraph
 or in math}}
\def\vnonvmode@#1#2{\relaxnext@\DNii@{\ifx\next\par\DN@{#1}\else
 \DN@{#2}\fi\next@}%
 \ifvmode\DN@{#1}\else
 \DN@{\FN@\nextii@}\fi\next@}
\def\newpage{\vnonvmode@{\vfill\break}{\nonvmodeerr@\newpage}}
\def\smallpagebreak{\vnonvmode@\smallbreak{\nonvmodeerr@\smallpagebreak}}
\def\medpagebreak{\vnonvmode@\medbreak{\nonvmodeerr@\medpagebreak}}
\def\bigpagebreak{\vnonvmode@\bigbreak{\nonvmodeerr@\bigpagebreak}}
\def\NoBlackBoxes{\global\overfullrule\z@}
\def\BlackBoxes{\global\overfullrule5\p@}
\def\Invalid@#1{\def#1{\Err@{\Invalid@@\string#1}}}
\def\Invalid@@{Invalid use of }
\message{figures,}
\Invalid@\caption
\Invalid@\captionwidth
\newdimen\smallcaptionwidth@
\def\topspace{\mid@false\ins@}
\def\midspace{\mid@true\ins@}
\newif\ifmid@
\def\captionfont@{}
\def\ins@#1{\relaxnext@\allowbreak
 \smallcaptionwidth@\captionwidth@\gdef\thespace@{#1}%
 \DN@{\ifx\next\space@\DN@. {\FN@\nextii@}\else
  \DN@.{\FN@\nextii@}\fi\next@.}%
 \DNii@{\ifx\next\caption\DN@\caption{\FN@\nextiii@}%
  \else\let\next@\nextiv@\fi\next@}%
 \def\nextiv@{\vnonvmode@
  {\ifmid@\expandafter\midinsert\else\expandafter\topinsert\fi
   \vbox to\thespace@{}\endinsert}
  {\ifmid@\nonvmodeerr@\midspace\else\nonvmodeerr@\topspace\fi}}%
 \def\nextiii@{\ifx\next\captionwidth\expandafter\nextv@
  \else\expandafter\nextvi@\fi}%
 \def\nextv@\captionwidth##1##2{\smallcaptionwidth@##1\relax\nextvi@{##2}}%
 \def\nextvi@##1{\def\thecaption@{\captionfont@##1}%
  \DN@{\ifx\next\space@\DN@. {\FN@\nextvii@}\else
   \DN@.{\FN@\nextvii@}\fi\next@.}%
  \FN@\next@}%
 \def\nextvii@{\vnonvmode@
  {\ifmid@\expandafter\midinsert\else
  \expandafter\topinsert\fi\vbox to\thespace@{}\nobreak\smallskip
  \setboxz@h{\noindent\ignorespaces\thecaption@\unskip}%
  \ifdim\wdz@>\smallcaptionwidth@\centerline{\vbox{\hsize\smallcaptionwidth@
   \noindent\ignorespaces\thecaption@\unskip}}%
  \else\centerline{\boxz@}\fi\endinsert}
  {\ifmid@\nonvmodeerr@\midspace
  \else\nonvmodeerr@\topspace\fi}}%
 \FN@\next@}
\message{comments,}
\def\newcodes@{\catcode`\\12\catcode`\{12\catcode`\}12\catcode`\#12%
 \catcode`\%12\relax}
\def\oldcodes@{\catcode`\\0\catcode`\{1\catcode`\}2\catcode`\#6%
 \catcode`\%14\relax}
\def\comment{\newcodes@\endlinechar=10 \comment@}
{\lccode`\0=`\\
\lowercase{\gdef\comment@#1^^J{\comment@@#10endcomment\comment@@@}%
\gdef\comment@@#10endcomment{\FN@\comment@@@}%
\gdef\comment@@@#1\comment@@@{\ifx\next\comment@@@\let\next\comment@
 \else\def\next{\oldcodes@\endlinechar=`\^^M\relax}%
 \fi\next}}}
\def\pr@m@s{\ifx'\next\DN@##1{\prim@s}\else\let\next@\egroup\fi\next@}
\def\prime{{\null\prime@\null}}
\mathchardef\prime@="0230
\let\dsize\displaystyle

\let\ssize\scriptstyle

\message{math spacing,}
\def\,{\RIfM@\mskip\thinmuskip\relax\else\kern.16667em\fi}
\def\!{\RIfM@\mskip-\thinmuskip\relax\else\kern-.16667em\fi}
\let\thinspace\,
\let\negthinspace\!
\def\medspace{\RIfM@\mskip\medmuskip\relax\else\kern.222222em\fi}
\def\negmedspace{\RIfM@\mskip-\medmuskip\relax\else\kern-.222222em\fi}
\def\thickspace{\RIfM@\mskip\thickmuskip\relax\else\kern.27777em\fi}
\let\;\thickspace
\def\negthickspace{\RIfM@\mskip-\thickmuskip\relax\else
 \kern-.27777em\fi}
\atdef@,{\RIfM@\mskip.1\thinmuskip\else\leavevmode\null,\fi}
\atdef@!{\RIfM@\mskip-.1\thinmuskip\else\leavevmode\null!\fi}
\atdef@.{\RIfM@&&\else\leavevmode.\spacefactor3000 \fi}
\def\and{\DOTSB\;\mathchar"3026 \;}

\message{fractions,}
\def\frac#1#2{{#1\over#2}}

\newdimen\ex@
\ex@.2326ex
\Invalid@\thickness
\def\thickfrac{\relaxnext@
 \DN@{\ifx\next\thickness\let\next@\nextii@\else
 \DN@{\nextii@\thickness1}\fi\next@}%
 \DNii@\thickness##1##2##3{{##2\above##1\ex@##3}}%
 \FN@\next@}

\def\thickfracwithdelims#1#2{\relaxnext@\def\ldelim@{#1}\def\rdelim@{#2}%
 \DN@{\ifx\next\thickness\let\next@\nextii@\else
 \DN@{\nextii@\thickness1}\fi\next@}%
 \DNii@\thickness##1##2##3{{##2\abovewithdelims
 \ldelim@\rdelim@##1\ex@##3}}%
 \FN@\next@}
\def\binom#1#2{{#1\choose#2}}

\def\:{\nobreak\hskip.1111em\mathpunct{}\nonscript\mkern-\thinmuskip{:}\hskip
 .3333emplus.0555em\relax}
\def\snug{\unskip\kern-\mathsurround}
\message{smash commands,}
\def\topsmash{\top@true\bot@false\smash@}
\def\botsmash{\top@false\bot@true\smash@}
\newif\iftop@
\newif\ifbot@
\def\smash{\top@true\bot@true\smash@}
\def\smash@{\RIfM@\expandafter\mathpalette\expandafter\mathsm@sh\else
 \expandafter\makesm@sh\fi}
\def\finsm@sh{\iftop@\ht\z@\z@\fi\ifbot@\dp\z@\z@\fi\leavevmode\boxz@}
\message{large operator symbols,}
\def\LimitsOnSums{\global\let\slimits@\displaylimits}
\def\NoLimitsOnSums{\global\let\slimits@\nolimits}
\LimitsOnSums
\mathchardef\coprod@="1360       \def\coprod{\DOTSB\coprod@\slimits@}
\mathchardef\bigvee@="1357       \def\bigvee{\DOTSB\bigvee@\slimits@}
\mathchardef\bigwedge@="1356     \def\bigwedge{\DOTSB\bigwedge@\slimits@}
\mathchardef\biguplus@="1355     \def\biguplus{\DOTSB\biguplus@\slimits@}
\mathchardef\bigcap@="1354       \def\bigcap{\DOTSB\bigcap@\slimits@}
\mathchardef\bigcup@="1353       \def\bigcup{\DOTSB\bigcup@\slimits@}
\mathchardef\prod@="1351         \def\prod{\DOTSB\prod@\slimits@}
\mathchardef\sum@="1350          \def\sum{\DOTSB\sum@\slimits@}
\mathchardef\bigotimes@="134E    \def\bigotimes{\DOTSB\bigotimes@\slimits@}
\mathchardef\bigoplus@="134C     \def\bigoplus{\DOTSB\bigoplus@\slimits@}
\mathchardef\bigodot@="134A      \def\bigodot{\DOTSB\bigodot@\slimits@}
\mathchardef\bigsqcup@="1346     \def\bigsqcup{\DOTSB\bigsqcup@\slimits@}
\message{integrals,}
\def\LimitsOnInts{\global\let\ilimits@\displaylimits}
\def\NoLimitsOnInts{\global\let\ilimits@\nolimits}
\NoLimitsOnInts
\def\int{\DOTSI\intop\ilimits@}
\def\oint{\DOTSI\ointop\ilimits@}
\def\intic@{\mathchoice{\hskip.5em}{\hskip.4em}{\hskip.4em}{\hskip.4em}}
\def\negintic@{\mathchoice
 {\hskip-.5em}{\hskip-.4em}{\hskip-.4em}{\hskip-.4em}}
\def\intkern@{\mathchoice{\!\!\!}{\!\!}{\!\!}{\!\!}}
\def\intdots@{\mathchoice{\plaincdots@}
 {{\cdotp}\mkern1.5mu{\cdotp}\mkern1.5mu{\cdotp}}
 {{\cdotp}\mkern1mu{\cdotp}\mkern1mu{\cdotp}}
 {{\cdotp}\mkern1mu{\cdotp}\mkern1mu{\cdotp}}}
\newcount\intno@
\def\iint{\DOTSI\intno@\tw@\FN@\ints@}
\def\iiint{\DOTSI\intno@\thr@@\FN@\ints@}
\def\iiiint{\DOTSI\intno@4 \FN@\ints@}
\def\idotsint{\DOTSI\intno@\z@\FN@\ints@}
\def\ints@{\findlimits@\ints@@}
\newif\iflimtoken@
\newif\iflimits@
\def\findlimits@{\limtoken@true\ifx\next\limits\limits@true
 \else\ifx\next\nolimits\limits@false\else
 \limtoken@false\ifx\ilimits@\nolimits\limits@false\else
 \ifinner\limits@false\else\limits@true\fi\fi\fi\fi}
\def\multint@{\int\ifnum\intno@=\z@\intdots@                                
 \else\intkern@\fi                                                          
 \ifnum\intno@>\tw@\int\intkern@\fi                                         
 \ifnum\intno@>\thr@@\int\intkern@\fi                                       
 \int}                                                                      
\def\multintlimits@{\intop\ifnum\intno@=\z@\intdots@\else\intkern@\fi
 \ifnum\intno@>\tw@\intop\intkern@\fi
 \ifnum\intno@>\thr@@\intop\intkern@\fi\intop}
\def\ints@@{\iflimtoken@                                                    
 \def\ints@@@{\iflimits@\negintic@\mathop{\intic@\multintlimits@}\limits    
  \else\multint@\nolimits\fi                                                
  \eat@}                                                                    
 \else                                                                      
 \def\ints@@@{\iflimits@\negintic@
  \mathop{\intic@\multintlimits@}\limits\else
  \multint@\nolimits\fi}\fi\ints@@@}
\def\LimitsOnNames{\global\let\nlimits@\displaylimits}
\def\NoLimitsOnNames{\global\let\nlimits@\nolimits@}
\LimitsOnNames
\def\nolimits@{\relaxnext@
 \DN@{\ifx\next\limits\DN@\limits{\nolimits}\else
  \let\next@\nolimits\fi\next@}%
 \FN@\next@}
\message{operator names,}
\def\newmcodes@{\mathcode`\'"27\mathcode`\*"2A\mathcode`\."613A%
 \mathcode`\-"2D\mathcode`\/"2F\mathcode`\:"603A }
\def\operatorname#1{\mathop{\newmcodes@\kern\z@\fam\z@#1}\nolimits@}
\def\operatornamewithlimits#1{\mathop{\newmcodes@\kern\z@\fam\z@#1}\nlimits@}
\def\qopname@#1{\mathop{\fam\z@#1}\nolimits@}
\def\qopnamewl@#1{\mathop{\fam\z@#1}\nlimits@}
\def\arccos{\qopname@{arccos}}
\def\arcsin{\qopname@{arcsin}}
\def\arctan{\qopname@{arctan}}
\def\arg{\qopname@{arg}}
\def\cos{\qopname@{cos}}
\def\cosh{\qopname@{cosh}}
\def\cot{\qopname@{cot}}
\def\coth{\qopname@{coth}}
\def\csc{\qopname@{csc}}
\def\deg{\qopname@{deg}}
\def\det{\qopnamewl@{det}}
\def\dim{\qopname@{dim}}
\def\exp{\qopname@{exp}}
\def\gcd{\qopnamewl@{gcd}}
\def\hom{\qopname@{hom}}
\def\inf{\qopnamewl@{inf}}
\def\injlim{\qopnamewl@{inj\,lim}}
\def\ker{\qopname@{ker}}
\def\lg{\qopname@{lg}}
\def\lim{\qopnamewl@{lim}}
\def\liminf{\qopnamewl@{lim\,inf}}
\def\limsup{\qopnamewl@{lim\,sup}}
\def\ln{\qopname@{ln}}
\def\log{\qopname@{log}}
\def\max{\qopnamewl@{max}}
\def\min{\qopnamewl@{min}}
\def\Pr{\qopnamewl@{Pr}}
\def\projlim{\qopnamewl@{proj\,lim}}
\def\sec{\qopname@{sec}}
\def\sin{\qopname@{sin}}
\def\sinh{\qopname@{sinh}}
\def\sup{\qopnamewl@{sup}}
\def\tan{\qopname@{tan}}
\def\tanh{\qopname@{tanh}}
\def\varinjlim{\mathop{\vtop{\ialign{##\crcr
 \hfil\rm lim\hfil\crcr\noalign{\nointerlineskip}\rightarrowfill\crcr
 \noalign{\nointerlineskip\kern-\ex@}\crcr}}}}
\def\varprojlim{\mathop{\vtop{\ialign{##\crcr
 \hfil\rm lim\hfil\crcr\noalign{\nointerlineskip}\leftarrowfill\crcr
 \noalign{\nointerlineskip\kern-\ex@}\crcr}}}}
\def\varliminf{\mathop{\underline{\vrule height\z@ depth.2exwidth\z@
 \hbox{\rm lim}}}}

\newdimen\buffer@
\buffer@\fontdimen13 \tenex
\newdimen\buffer
\buffer\buffer@

\def\ResetBuffer{\fontdimen13 \tenex\buffer@\global\buffer\buffer@}
\def\shave#1{\mathop{\hbox{$\m@th\fontdimen13 \tenex\z@                     
 \displaystyle{#1}$}}\fontdimen13 \tenex\buffer}

\message{multilevel sub/superscripts,}
\Invalid@\\
\def\Let@{\relax\iffalse{\fi\let\\=\cr\iffalse}\fi}
\Invalid@\vspace
\def\vspace@{\def\vspace##1{\crcr\noalign{\vskip##1\relax}}}
\def\multilimits@{\bgroup\vspace@\Let@
 \baselineskip\fontdimen10 \scriptfont\tw@
 \advance\baselineskip\fontdimen12 \scriptfont\tw@
 \lineskip\thr@@\fontdimen8 \scriptfont\thr@@
 \lineskiplimit\lineskip
 \vbox\bgroup\ialign\bgroup\hfil$\m@th\scriptstyle{##}$\hfil\crcr}
\def\Sb{_\multilimits@}
\def\endSb{\crcr\egroup\egroup\egroup}
\def\Sp{^\multilimits@}

\def\spreadlines#1{\RIfMIfI@\onlydmatherr@\spreadlines\else
 \openup#1\relax\fi\else\onlydmatherr@\spreadlines\fi}
\def\Mathstrut@{\copy\Mathstrutbox@}
\newbox\Mathstrutbox@
\setbox\Mathstrutbox@\null
\setboxz@h{$\m@th($}
\ht\Mathstrutbox@\ht\z@
\dp\Mathstrutbox@\dp\z@
\message{matrices,}
\newdimen\spreadmlines@
\def\spreadmatrixlines#1{\RIfMIfI@
 \onlydmatherr@\spreadmatrixlines\else
 \spreadmlines@#1\relax\fi\else\onlydmatherr@\spreadmatrixlines\fi}
\def\format{\crcr\egroup\iffalse{\fi\ifnum`}=0 \fi\format@}
\newtoks\hashtoks@
\hashtoks@{#}
\def\format@#1\\{\def\preamble@{#1}%
 \def\l{$\m@th\the\hashtoks@$\hfil}%
 \def\c{\hfil$\m@th\the\hashtoks@$\hfil}%
 \def\r{\hfil$\m@th\the\hashtoks@$}%
 \edef\preamble@@{\preamble@}\ifnum`{=0 \fi\iffalse}\fi
 \ialign\bgroup\span\preamble@@\crcr}
\def\smallmatrix{\null\,\vcenter\bgroup\vspace@\Let@
 \baselineskip9\ex@\lineskip\ex@
 \ialign\bgroup\hfil$\m@th\scriptstyle{##}$\hfil&&\thickspace\hfil
 $\m@th\scriptstyle{##}$\hfil\crcr}
\def\endsmallmatrix{\crcr\egroup\egroup\,}

\newmuskip\dotsspace@
\dotsspace@1.5mu
\def\strip@#1 {#1}
\def\spacehdots#1\for#2{\multispan{#2}\xleaders
 \hbox{$\m@th\mkern\strip@#1 \dotsspace@.\mkern\strip@#1 \dotsspace@$}\hfill}
\def\hdotsfor#1{\spacehdots\@ne\for{#1}}
\def\multispan@#1{\omit\mscount#1\unskip\loop\ifnum\mscount>\@ne\sp@n\repeat}
\def\spaceinnerhdots#1\for#2\after#3{\multispan@{\strip@#2 }#3\xleaders
 \hbox{$\m@th\mkern\strip@#1 \dotsspace@.\mkern\strip@#1 \dotsspace@$}\hfill}
\def\innerhdotsfor#1\after#2{\spaceinnerhdots\@ne\for#1\after{#2}}
\def\endcases{\endmatrix\right.\egroup}
\message{multiline displays,}
\newif\ifinany@
\newif\ifinalign@
\newif\ifingather@
\def\strut@{\copy\strutbox@}
\newbox\strutbox@
\setbox\strutbox@\hbox{\vrule height8\p@ depth3\p@ width\z@}
\def\topaligned{\null\,\vtop\aligned@}
\def\botaligned{\null\,\vbox\aligned@}
\def\aligned{\null\,\vcenter\aligned@}
\def\aligned@{\bgroup\vspace@\Let@
 \ifinany@\else\openup\jot\fi\ialign
 \bgroup\hfil\strut@$\m@th\displaystyle{##}$&
 $\m@th\displaystyle{{}##}$\hfil\crcr}
\def\endaligned{\crcr\egroup\egroup}

\def\alignedat#1{\null\,\vcenter\bgroup\doat@{#1}\vspace@\Let@
 \ifinany@\else\openup\jot\fi\ialign\bgroup\span\preamble@@\crcr}
\newcount\atcount@
\def\doat@#1{\toks@{\hfil\strut@$\m@th
 \displaystyle{\the\hashtoks@}$&$\m@th\displaystyle
 {{}\the\hashtoks@}$\hfil}
 \atcount@#1\relax\advance\atcount@\m@ne                                    
 \loop\ifnum\atcount@>\z@\toks@=\expandafter{\the\toks@&\hfil$\m@th
 \displaystyle{\the\hashtoks@}$&$\m@th
 \displaystyle{{}\the\hashtoks@}$\hfil}\advance
  \atcount@\m@ne\repeat                                                     
 \xdef\preamble@{\the\toks@}\xdef\preamble@@{\preamble@}}

\def\gathered{\null\,\vcenter\bgroup\vspace@\Let@
 \ifinany@\else\openup\jot\fi\ialign
 \bgroup\hfil\strut@$\m@th\displaystyle{##}$\hfil\crcr}
\def\endgathered{\crcr\egroup\egroup}
\newif\iftagsleft@
\def\TagsOnLeft{\global\tagsleft@true}
\def\TagsOnRight{\global\tagsleft@false}
\TagsOnLeft
\newif\ifmathtags@
\def\TagsAsMath{\global\mathtags@true}
\def\TagsAsText{\global\mathtags@false}
\TagsAsText
\def\tagform@#1{\hbox{\rm(\ignorespaces#1\unskip)}}
\def\thetag{\leavevmode\tagform@}
\def\tag#1$${\iftagsleft@\leqno\else\eqno\fi                                
 \maketag@#1\maketag@                                                       
 $$}                                                                        
\def\maketag@{\FN@\maketag@@}
\def\maketag@@{\ifx\next"\expandafter\maketag@@@\else\expandafter\maketag@@@@
 \fi}
\def\maketag@@@"#1"#2\maketag@{\hbox{\rm#1}}                                
\def\maketag@@@@#1\maketag@{\ifmathtags@\tagform@{$\m@th#1$}\else
 \tagform@{#1}\fi}
\interdisplaylinepenalty\@M
\def\allowdisplaybreaks{\RIfMIfI@
 \onlydmatherr@\allowdisplaybreaks\else
 \interdisplaylinepenalty\z@\fi\else\onlydmatherr@\allowdisplaybreaks\fi}
\Invalid@\allowdisplaybreak
\Invalid@\displaybreak
\Invalid@\intertext
\def\allowdisplaybreak@{\def\allowdisplaybreak{\crcr\noalign{\allowbreak}}}
\def\displaybreak@{\def\displaybreak{\crcr\noalign{\break}}}
\def\intertext@{\def\intertext##1{\crcr\noalign{%
 \penalty\postdisplaypenalty \vskip\belowdisplayskip
 \vbox{\normalbaselines\noindent##1}%
 \penalty\predisplaypenalty \vskip\abovedisplayskip}}}
\newskip\centering@
\centering@\z@ plus\@m\p@
\def\align{\relax\ifingather@\DN@{\csname align (in
  \string\gather)\endcsname}\else
 \ifmmode\ifinner\DN@{\onlydmatherr@\align}\else
  \let\next@\align@\fi
 \else\DN@{\onlydmatherr@\align}\fi\fi\next@}
\newhelp\andhelp@
{An extra & here is so disastrous that you should probably exit^^J
and fix things up.}
\newif\iftag@
\newcount\and@
\def\align@{\inalign@true\inany@true
 \vspace@\allowdisplaybreak@\displaybreak@\intertext@
 \def\tag{\global\tag@true\ifnum\and@=\z@\DN@{&&}\else
          \DN@{&}\fi\next@}%
 \iftagsleft@\DN@{\csname align \endcsname}\else
  \DN@{\csname align \space\endcsname}\fi\next@}
\def\Tag@{\iftag@\else\errhelp\andhelp@\err@{Extra & on this line}\fi}
\newdimen\lwidth@
\newdimen\rwidth@
\newdimen\maxlwidth@
\newdimen\maxrwidth@
\newdimen\totwidth@
\def\measure@#1\endalign{\lwidth@\z@\rwidth@\z@\maxlwidth@\z@\maxrwidth@\z@
 \global\and@\z@                                                            
 \setbox@ne\vbox                                                            
  {\everycr{\noalign{\global\tag@false\global\and@\z@}}\Let@                
  \halign{\setboxz@h{$\m@th\displaystyle{\@lign##}$}
   \global\lwidth@\wdz@                                                     
   \ifdim\lwidth@>\maxlwidth@\global\maxlwidth@\lwidth@\fi                  
   \global\advance\and@\@ne                                                 
   &\setboxz@h{$\m@th\displaystyle{{}\@lign##}$}\global\rwidth@\wdz@        
   \ifdim\rwidth@>\maxrwidth@\global\maxrwidth@\rwidth@\fi                  
   \global\advance\and@\@ne                                                
   &\Tag@
   \eat@{##}\crcr#1\crcr}}
 \totwidth@\maxlwidth@\advance\totwidth@\maxrwidth@}                       
\def\displ@y@{\global\dt@ptrue\openup\jot
 \everycr{\noalign{\global\tag@false\global\and@\z@\ifdt@p\global\dt@pfalse
 \vskip-\lineskiplimit\vskip\normallineskiplimit\else
 \penalty\interdisplaylinepenalty\fi}}}
\def\black@#1{\noalign{\ifdim#1>\displaywidth
 \dimen@\prevdepth\nointerlineskip                                          
 \vskip-\ht\strutbox@\vskip-\dp\strutbox@                                   
 \vbox{\noindent\hbox to#1{\strut@\hfill}}
 \prevdepth\dimen@                                                          
 \fi}}
\expandafter\def\csname align \space\endcsname#1\endalign
 {\measure@#1\endalign\global\and@\z@                                       
 \ifingather@\everycr{\noalign{\global\and@\z@}}\else\displ@y@\fi           
 \Let@\tabskip\centering@                                                   
 \halign to\displaywidth
  {\hfil\strut@\setboxz@h{$\m@th\displaystyle{\@lign##}$}
  \global\lwidth@\wdz@\boxz@\global\advance\and@\@ne                        
  \tabskip\z@skip                                                           
  &\setboxz@h{$\m@th\displaystyle{{}\@lign##}$}
  \global\rwidth@\wdz@\boxz@\hfill\global\advance\and@\@ne                  
  \tabskip\centering@                                                       
  &\setboxz@h{\@lign\strut@\maketag@##\maketag@}
  \dimen@\displaywidth\advance\dimen@-\totwidth@
  \divide\dimen@\tw@\advance\dimen@\maxrwidth@\advance\dimen@-\rwidth@     
  \ifdim\dimen@<\tw@\wdz@\llap{\vtop{\normalbaselines\null\boxz@}}
  \else\llap{\boxz@}\fi                                                    
  \tabskip\z@skip                                                          
  \crcr#1\crcr                                                             
  \black@\totwidth@}}                                                      
\newdimen\lineht@
\expandafter\def\csname align \endcsname#1\endalign{\measure@#1\endalign
 \global\and@\z@
 \ifdim\totwidth@>\displaywidth\let\displaywidth@\totwidth@\else
  \let\displaywidth@\displaywidth\fi                                        
 \ifingather@\everycr{\noalign{\global\and@\z@}}\else\displ@y@\fi
 \Let@\tabskip\centering@\halign to\displaywidth
  {\hfil\strut@\setboxz@h{$\m@th\displaystyle{\@lign##}$}%
  \global\lwidth@\wdz@\global\lineht@\ht\z@                                 
  \boxz@\global\advance\and@\@ne
  \tabskip\z@skip&\setboxz@h{$\m@th\displaystyle{{}\@lign##}$}%
  \global\rwidth@\wdz@\ifdim\ht\z@>\lineht@\global\lineht@\ht\z@\fi         
  \boxz@\hfil\global\advance\and@\@ne
  \tabskip\centering@&\kern-\displaywidth@                                  
  \setboxz@h{\@lign\strut@\maketag@##\maketag@}%
  \dimen@\displaywidth\advance\dimen@-\totwidth@
  \divide\dimen@\tw@\advance\dimen@\maxlwidth@\advance\dimen@-\lwidth@
  \ifdim\dimen@<\tw@\wdz@
   \rlap{\vbox{\normalbaselines\boxz@\vbox to\lineht@{}}}\else
   \rlap{\boxz@}\fi
  \tabskip\displaywidth@\crcr#1\crcr\black@\totwidth@}}
\expandafter\def\csname align (in \string\gather)\endcsname
  #1\endalign{\vcenter{\align@#1\endalign}}
\Invalid@\endalign
\newif\ifxat@
\def\alignat{\RIfMIfI@\DN@{\onlydmatherr@\alignat}\else
 \DN@{\csname alignat \endcsname}\fi\else
 \DN@{\onlydmatherr@\alignat}\fi\next@}
\newif\ifmeasuring@
\newbox\savealignat@
\expandafter\def\csname alignat \endcsname#1#2\endalignat                   
 {\inany@true\xat@false
 \def\tag{\global\tag@true\count@#1\relax\multiply\count@\tw@
  \xdef\tag@{}\loop\ifnum\count@>\and@\xdef\tag@{&\tag@}\advance\count@\m@ne
  \repeat\tag@}%
 \vspace@\allowdisplaybreak@\displaybreak@\intertext@
 \displ@y@\measuring@true                                                   
 \setbox\savealignat@\hbox{$\m@th\displaystyle\Let@
  \attag@{#1}
  \vbox{\halign{\span\preamble@@\crcr#2\crcr}}$}%
 \measuring@false                                                           
 \Let@\attag@{#1}
 \tabskip\centering@\halign to\displaywidth
  {\span\preamble@@\crcr#2\crcr                                             
  \black@{\wd\savealignat@}}}                                               
\Invalid@\endalignat
\def\xalignat{\RIfMIfI@
 \DN@{\onlydmatherr@\xalignat}\else
 \DN@{\csname xalignat \endcsname}\fi\else
 \DN@{\onlydmatherr@\xalignat}\fi\next@}
\expandafter\def\csname xalignat \endcsname#1#2\endxalignat
 {\inany@true\xat@true
 \def\tag{\global\tag@true\def\tag@{}\count@#1\relax\multiply\count@\tw@
  \loop\ifnum\count@>\and@\xdef\tag@{&\tag@}\advance\count@\m@ne\repeat\tag@}%
 \vspace@\allowdisplaybreak@\displaybreak@\intertext@
 \displ@y@\measuring@true\setbox\savealignat@\hbox{$\m@th\displaystyle\Let@
 \attag@{#1}\vbox{\halign{\span\preamble@@\crcr#2\crcr}}$}%
 \measuring@false\Let@
 \attag@{#1}\tabskip\centering@\halign to\displaywidth
 {\span\preamble@@\crcr#2\crcr\black@{\wd\savealignat@}}}
\def\attag@#1{\let\Maketag@\maketag@\let\TAG@\Tag@                          
 \let\Tag@=0\let\maketag@=0
 \ifmeasuring@\def\llap@##1{\setboxz@h{##1}\hbox to\tw@\wdz@{}}%
  \def\rlap@##1{\setboxz@h{##1}\hbox to\tw@\wdz@{}}\else
  \let\llap@\llap\let\rlap@\rlap\fi                                         
 \toks@{\hfil\strut@$\m@th\displaystyle{\@lign\the\hashtoks@}$\tabskip\z@skip
  \global\advance\and@\@ne&$\m@th\displaystyle{{}\@lign\the\hashtoks@}$\hfil
  \ifxat@\tabskip\centering@\fi\global\advance\and@\@ne}
 \iftagsleft@
  \toks@@{\tabskip\centering@&\Tag@\kern-\displaywidth
   \rlap@{\@lign\maketag@\the\hashtoks@\maketag@}%
   \global\advance\and@\@ne\tabskip\displaywidth}\else
  \toks@@{\tabskip\centering@&\Tag@\llap@{\@lign\maketag@
   \the\hashtoks@\maketag@}\global\advance\and@\@ne\tabskip\z@skip}\fi      
 \atcount@#1\relax\advance\atcount@\m@ne
 \loop\ifnum\atcount@>\z@
 \toks@=\expandafter{\the\toks@&\hfil$\m@th\displaystyle{\@lign
  \the\hashtoks@}$\global\advance\and@\@ne
  \tabskip\z@skip&$\m@th\displaystyle{{}\@lign\the\hashtoks@}$\hfil\ifxat@
  \tabskip\centering@\fi\global\advance\and@\@ne}\advance\atcount@\m@ne
 \repeat                                                                    
 \xdef\preamble@{\the\toks@\the\toks@@}
 \xdef\preamble@@{\preamble@}
 \let\maketag@\Maketag@\let\Tag@\TAG@}                                      
\Invalid@\endxalignat
\def\xxalignat{\RIfMIfI@
 \DN@{\onlydmatherr@\xxalignat}\else\DN@{\csname xxalignat
  \endcsname}\fi\else
 \DN@{\onlydmatherr@\xxalignat}\fi\next@}
\expandafter\def\csname xxalignat \endcsname#1#2\endxxalignat{\inany@true
 \vspace@\allowdisplaybreak@\displaybreak@\intertext@
 \displ@y\setbox\savealignat@\hbox{$\m@th\displaystyle\Let@
 \xxattag@{#1}\vbox{\halign{\span\preamble@@\crcr#2\crcr}}$}%
 \Let@\xxattag@{#1}\tabskip\z@skip\halign to\displaywidth
 {\span\preamble@@\crcr#2\crcr\black@{\wd\savealignat@}}}
\def\xxattag@#1{\toks@{\tabskip\z@skip\hfil\strut@
 $\m@th\displaystyle{\the\hashtoks@}$&%
 $\m@th\displaystyle{{}\the\hashtoks@}$\hfil\tabskip\centering@&}%
 \atcount@#1\relax\advance\atcount@\m@ne\loop\ifnum\atcount@>\z@
 \toks@=\expandafter{\the\toks@&\hfil$\m@th\displaystyle{\the\hashtoks@}$%
  \tabskip\z@skip&$\m@th\displaystyle{{}\the\hashtoks@}$\hfil
  \tabskip\centering@}\advance\atcount@\m@ne\repeat
 \xdef\preamble@{\the\toks@\tabskip\z@skip}\xdef\preamble@@{\preamble@}}
\Invalid@\endxxalignat
\newdimen\gwidth@
\newdimen\gmaxwidth@
\def\gmeasure@#1\endgather{\gwidth@\z@\gmaxwidth@\z@\setbox@ne\vbox{\Let@
 \halign{\setboxz@h{$\m@th\displaystyle{##}$}\global\gwidth@\wdz@
 \ifdim\gwidth@>\gmaxwidth@\global\gmaxwidth@\gwidth@\fi
 &\eat@{##}\crcr#1\crcr}}}
\def\gather{\RIfMIfI@\DN@{\onlydmatherr@\gather}\else
 \ingather@true\inany@true\def\tag{&}%
 \vspace@\allowdisplaybreak@\displaybreak@\intertext@
 \displ@y\Let@
 \iftagsleft@\DN@{\csname gather \endcsname}\else
  \DN@{\csname gather \space\endcsname}\fi\fi
 \else\DN@{\onlydmatherr@\gather}\fi\next@}
\expandafter\def\csname gather \space\endcsname#1\endgather
 {\gmeasure@#1\endgather\tabskip\centering@
 \halign to\displaywidth{\hfil\strut@\setboxz@h{$\m@th\displaystyle{##}$}%
 \global\gwidth@\wdz@\boxz@\hfil&
 \setboxz@h{\strut@{\maketag@##\maketag@}}%
 \dimen@\displaywidth\advance\dimen@-\gwidth@
 \ifdim\dimen@>\tw@\wdz@\llap{\boxz@}\else
 \llap{\vtop{\normalbaselines\null\boxz@}}\fi
 \tabskip\z@skip\crcr#1\crcr\black@\gmaxwidth@}}
\newdimen\glineht@
\expandafter\def\csname gather \endcsname#1\endgather{\gmeasure@#1\endgather
 \ifdim\gmaxwidth@>\displaywidth\let\gdisplaywidth@\gmaxwidth@\else
 \let\gdisplaywidth@\displaywidth\fi\tabskip\centering@\halign to\displaywidth
 {\hfil\strut@\setboxz@h{$\m@th\displaystyle{##}$}%
 \global\gwidth@\wdz@\global\glineht@\ht\z@\boxz@\hfil&\kern-\gdisplaywidth@
 \setboxz@h{\strut@{\maketag@##\maketag@}}%
 \dimen@\displaywidth\advance\dimen@-\gwidth@
 \ifdim\dimen@>\tw@\wdz@\rlap{\boxz@}\else
 \rlap{\vbox{\normalbaselines\boxz@\vbox to\glineht@{}}}\fi
 \tabskip\gdisplaywidth@\crcr#1\crcr\black@\gmaxwidth@}}
\newif\ifctagsplit@
\def\CenteredTagsOnSplits{\global\ctagsplit@true}
\def\TopOrBottomTagsOnSplits{\global\ctagsplit@false}
\TopOrBottomTagsOnSplits
\def\split{\relax\ifinany@\let\next@\insplit@\else
 \ifmmode\ifinner\def\next@{\onlydmatherr@\split}\else
 \let\next@\outsplit@\fi\else
 \def\next@{\onlydmatherr@\split}\fi\fi\next@}
\def\insplit@{\global\setbox\z@\vbox\bgroup\vspace@\Let@\ialign\bgroup
 \hfil\strut@$\m@th\displaystyle{##}$&$\m@th\displaystyle{{}##}$\hfill\crcr}
\def\endsplit{\crcr\egroup\egroup\iftagsleft@\expandafter\lendsplit@\else
 \expandafter\rendsplit@\fi}
\def\rendsplit@{\global\setbox9 \vbox
 {\unvcopy\z@\global\setbox8 \lastbox\unskip}
 \setbox@ne\hbox{\unhcopy8 \unskip\global\setbox\tw@\lastbox
 \unskip\global\setbox\thr@@\lastbox}
 \global\setbox7 \hbox{\unhbox\tw@\unskip}
 \ifinalign@\ifctagsplit@                                                   
  \gdef\split@{\hbox to\wd\thr@@{}&
   \vcenter{\vbox{\moveleft\wd\thr@@\boxz@}}}
 \else\gdef\split@{&\vbox{\moveleft\wd\thr@@\box9}\crcr
  \box\thr@@&\box7}\fi                                                      
 \else                                                                      
  \ifctagsplit@\gdef\split@{\vcenter{\boxz@}}\else
  \gdef\split@{\box9\crcr\hbox{\box\thr@@\box7}}\fi
 \fi
 \split@}                                                                   
\def\lendsplit@{\global\setbox9\vtop{\unvcopy\z@}
 \setbox@ne\vbox{\unvcopy\z@\global\setbox8\lastbox}
 \setbox@ne\hbox{\unhcopy8\unskip\setbox\tw@\lastbox
  \unskip\global\setbox\thr@@\lastbox}
 \ifinalign@\ifctagsplit@                                                   
  \gdef\split@{\hbox to\wd\thr@@{}&
  \vcenter{\vbox{\moveleft\wd\thr@@\box9}}}
  \else                                                                     
  \gdef\split@{\hbox to\wd\thr@@{}&\vbox{\moveleft\wd\thr@@\box9}}\fi
 \else
  \ifctagsplit@\gdef\split@{\vcenter{\box9}}\else
  \gdef\split@{\box9}\fi
 \fi\split@}
\def\outsplit@#1$${\align\insplit@#1\endalign$$}
\newdimen\multlinegap@
\multlinegap@1em
\newdimen\multlinetaggap@
\multlinetaggap@1em
\def\MultlineGap#1{\global\multlinegap@#1\relax}
\def\multlinegap#1{\RIfMIfI@\onlydmatherr@\multlinegap\else
 \multlinegap@#1\relax\fi\else\onlydmatherr@\multlinegap\fi}
\def\nomultlinegap{\multlinegap{\z@}}
\def\multline{\RIfMIfI@
 \DN@{\onlydmatherr@\multline}\else
 \DN@{\multline@}\fi\else
 \DN@{\onlydmatherr@\multline}\fi\next@}
\newif\iftagin@
\def\tagin@#1{\tagin@false\in@\tag{#1}\ifin@\tagin@true\fi}
\def\multline@#1$${\inany@true\vspace@\allowdisplaybreak@\displaybreak@
 \tagin@{#1}\iftagsleft@\DN@{\multline@l#1$$}\else
 \DN@{\multline@r#1$$}\fi\next@}
\newdimen\mwidth@
\def\rmmeasure@#1\endmultline{%
 \def\shoveleft##1{##1}\def\shoveright##1{##1}
 \setbox@ne\vbox{\Let@\halign{\setboxz@h
  {$\m@th\@lign\displaystyle{}##$}\global\mwidth@\wdz@
  \crcr#1\crcr}}}
\newdimen\mlineht@
\newif\ifzerocr@
\newif\ifonecr@
\def\lmmeasure@#1\endmultline{\global\zerocr@true\global\onecr@false
 \everycr{\noalign{\ifonecr@\global\onecr@false\fi
  \ifzerocr@\global\zerocr@false\global\onecr@true\fi}}
  \def\shoveleft##1{##1}\def\shoveright##1{##1}%
 \setbox@ne\vbox{\Let@\halign{\setboxz@h
  {$\m@th\@lign\displaystyle{}##$}\ifonecr@\global\mwidth@\wdz@
  \global\mlineht@\ht\z@\fi\crcr#1\crcr}}}
\newbox\mtagbox@
\newdimen\ltwidth@
\newdimen\rtwidth@
\def\multline@l#1$${\iftagin@\DN@{\lmultline@@#1$$}\else
 \DN@{\setbox\mtagbox@\null\ltwidth@\z@\rtwidth@\z@
  \lmultline@@@#1$$}\fi\next@}
\def\lmultline@@#1\endmultline\tag#2$${%
 \setbox\mtagbox@\hbox{\maketag@#2\maketag@}
 \lmmeasure@#1\endmultline\dimen@\mwidth@\advance\dimen@\wd\mtagbox@
 \advance\dimen@\multlinetaggap@                                            
 \ifdim\dimen@>\displaywidth\ltwidth@\z@\else\ltwidth@\wd\mtagbox@\fi       
 \lmultline@@@#1\endmultline$$}
\def\lmultline@@@{\displ@y
 \def\shoveright##1{##1\hfilneg\hskip\multlinegap@}%
 \def\shoveleft##1{\setboxz@h{$\m@th\displaystyle{}##1$}%
  \setbox@ne\hbox{$\m@th\displaystyle##1$}%
  \hfilneg
  \iftagin@
   \ifdim\ltwidth@>\z@\hskip\ltwidth@\hskip\multlinetaggap@\fi
  \else\hskip\multlinegap@\fi\hskip.5\wd@ne\hskip-.5\wdz@##1}
  \halign\bgroup\Let@\hbox to\displaywidth
   {\strut@$\m@th\displaystyle\hfil{}##\hfil$}\crcr
   \hfilneg                                                                 
   \iftagin@                                                                
    \ifdim\ltwidth@>\z@                                                     
     \box\mtagbox@\hskip\multlinetaggap@                                    
    \else
     \rlap{\vbox{\normalbaselines\hbox{\strut@\box\mtagbox@}%
     \vbox to\mlineht@{}}}\fi                                               
   \else\hskip\multlinegap@\fi}                                             
\def\multline@r#1$${\iftagin@\DN@{\rmultline@@#1$$}\else
 \DN@{\setbox\mtagbox@\null\ltwidth@\z@\rtwidth@\z@
  \rmultline@@@#1$$}\fi\next@}
\def\rmultline@@#1\endmultline\tag#2$${\ltwidth@\z@
 \setbox\mtagbox@\hbox{\maketag@#2\maketag@}%
 \rmmeasure@#1\endmultline\dimen@\mwidth@\advance\dimen@\wd\mtagbox@
 \advance\dimen@\multlinetaggap@
 \ifdim\dimen@>\displaywidth\rtwidth@\z@\else\rtwidth@\wd\mtagbox@\fi
 \rmultline@@@#1\endmultline$$}
\def\rmultline@@@{\displ@y
 \def\shoveright##1{##1\hfilneg\iftagin@\ifdim\rtwidth@>\z@
  \hskip\rtwidth@\hskip\multlinetaggap@\fi\else\hskip\multlinegap@\fi}%
 \def\shoveleft##1{\setboxz@h{$\m@th\displaystyle{}##1$}%
  \setbox@ne\hbox{$\m@th\displaystyle##1$}%
  \hfilneg\hskip\multlinegap@\hskip.5\wd@ne\hskip-.5\wdz@##1}%
 \halign\bgroup\Let@\hbox to\displaywidth
  {\strut@$\m@th\displaystyle\hfil{}##\hfil$}\crcr
 \hfilneg\hskip\multlinegap@}
\def\endmultline{\iftagsleft@\expandafter\lendmultline@\else
 \expandafter\rendmultline@\fi}
\def\lendmultline@{\hfilneg\hskip\multlinegap@\crcr\egroup}
\def\rendmultline@{\iftagin@                                                
 \ifdim\rtwidth@>\z@                                                        
  \hskip\multlinetaggap@\box\mtagbox@                                       
 \else\llap{\vtop{\normalbaselines\null\hbox{\strut@\box\mtagbox@}}}\fi     
 \else\hskip\multlinegap@\fi                                                
 \hfilneg\crcr\egroup}
\def\bmod{\mskip-\medmuskip\mkern5mu\mathbin{\fam\z@ mod}\penalty900
 \mkern5mu\mskip-\medmuskip}
\def\pmod#1{\allowbreak\ifinner\mkern8mu\else\mkern18mu\fi
 ({\fam\z@ mod}\,\,#1)}
\def\pod#1{\allowbreak\ifinner\mkern8mu\else\mkern18mu\fi(#1)}
\def\mod#1{\allowbreak\ifinner\mkern12mu\else\mkern18mu\fi{\fam\z@ mod}\,\,#1}
\message{continued fractions,}
\newcount\cfraccount@
\def\cfrac{\bgroup\bgroup\advance\cfraccount@\@ne\strut
 \iffalse{\fi\def\\{\over\displaystyle}\iffalse}\fi}
\def\lcfrac{\bgroup\bgroup\advance\cfraccount@\@ne\strut
 \iffalse{\fi\def\\{\hfill\over\displaystyle}\iffalse}\fi}
\def\rcfrac{\bgroup\bgroup\advance\cfraccount@\@ne\strut\hfill
 \iffalse{\fi\def\\{\over\displaystyle}\iffalse}\fi}
\def\gloop@#1\repeat{\gdef\body{#1}\iterate}
\def\endcfrac{\gloop@\ifnum\cfraccount@>\z@\global\advance\cfraccount@\m@ne
 \egroup\hskip-\nulldelimiterspace\egroup\repeat}
\message{compound symbols,}
\def\binrel@#1{\setboxz@h{\thinmuskip0mu
  \medmuskip\m@ne mu\thickmuskip\@ne mu$#1\m@th$}%
 \setbox@ne\hbox{\thinmuskip0mu\medmuskip\m@ne mu\thickmuskip
  \@ne mu${}#1{}\m@th$}%
 \setbox\tw@\hbox{\hskip\wd@ne\hskip-\wdz@}}
\def\overset#1\to#2{\binrel@{#2}\ifdim\wd\tw@<\z@
 \mathbin{\mathop{\kern\z@#2}\limits^{#1}}\else\ifdim\wd\tw@>\z@
 \mathrel{\mathop{\kern\z@#2}\limits^{#1}}\else
 {\mathop{\kern\z@#2}\limits^{#1}}{}\fi\fi}
\def\underset#1\to#2{\binrel@{#2}\ifdim\wd\tw@<\z@
 \mathbin{\mathop{\kern\z@#2}\limits_{#1}}\else\ifdim\wd\tw@>\z@
 \mathrel{\mathop{\kern\z@#2}\limits_{#1}}\else
 {\mathop{\kern\z@#2}\limits_{#1}}{}\fi\fi}
\def\oversetbrace#1\to#2{\overbrace{#2}^{#1}}
\def\undersetbrace#1\to#2{\underbrace{#2}_{#1}}
\def\sideset#1\and#2\to#3{%
 \setbox@ne\hbox{$\dsize{\vphantom{#3}}#1{#3}\m@th$}%
 \setbox\tw@\hbox{$\dsize{#3}#2\m@th$}%
 \hskip\wd@ne\hskip-\wd\tw@\mathop{\hskip\wd\tw@\hskip-\wd@ne
  {\vphantom{#3}}#1{#3}#2}}
\def\rightarrowfill@#1{\setboxz@h{$#1-\m@th$}\ht\z@\z@
  $#1\m@th\copy\z@\mkern-6mu\cleaders
  \hbox{$#1\mkern-2mu\box\z@\mkern-2mu$}\hfill
  \mkern-6mu\mathord\rightarrow$}
\def\leftarrowfill@#1{\setboxz@h{$#1-\m@th$}\ht\z@\z@
  $#1\m@th\mathord\leftarrow\mkern-6mu\cleaders
  \hbox{$#1\mkern-2mu\copy\z@\mkern-2mu$}\hfill
  \mkern-6mu\box\z@$}
\def\leftrightarrowfill@#1{\setboxz@h{$#1-\m@th$}\ht\z@\z@
  $#1\m@th\mathord\leftarrow\mkern-6mu\cleaders
  \hbox{$#1\mkern-2mu\box\z@\mkern-2mu$}\hfill
  \mkern-6mu\mathord\rightarrow$}
\def\overrightarrow{\mathpalette\overrightarrow@}
\def\overrightarrow@#1#2{\vbox{\ialign{##\crcr\rightarrowfill@#1\crcr
 \noalign{\kern-\ex@\nointerlineskip}$\m@th\hfil#1#2\hfil$\crcr}}}

\def\overleftarrow{\mathpalette\overleftarrow@}
\def\overleftarrow@#1#2{\vbox{\ialign{##\crcr\leftarrowfill@#1\crcr
 \noalign{\kern-\ex@\nointerlineskip}$\m@th\hfil#1#2\hfil$\crcr}}}
\def\overleftrightarrow{\mathpalette\overleftrightarrow@}
\def\overleftrightarrow@#1#2{\vbox{\ialign{##\crcr\leftrightarrowfill@#1\crcr
 \noalign{\kern-\ex@\nointerlineskip}$\m@th\hfil#1#2\hfil$\crcr}}}
\def\underrightarrow{\mathpalette\underrightarrow@}
\def\underrightarrow@#1#2{\vtop{\ialign{##\crcr$\m@th\hfil#1#2\hfil$\crcr
 \noalign{\nointerlineskip}\rightarrowfill@#1\crcr}}}

\def\underleftarrow{\mathpalette\underleftarrow@}
\def\underleftarrow@#1#2{\vtop{\ialign{##\crcr$\m@th\hfil#1#2\hfil$\crcr
 \noalign{\nointerlineskip}\leftarrowfill@#1\crcr}}}
\def\underleftrightarrow{\mathpalette\underleftrightarrow@}
\def\underleftrightarrow@#1#2{\vtop{\ialign{##\crcr$\m@th\hfil#1#2\hfil$\crcr
 \noalign{\nointerlineskip}\leftrightarrowfill@#1\crcr}}}
\message{various kinds of dots,}
\let\DOTSI\relax
\let\DOTSB\relax

\newif\ifmath@
{\uccode`7=`\\ \uccode`8=`m \uccode`9=`a \uccode`0=`t \uccode`!=`h
 \uppercase{\gdef\math@#1#2#3#4#5#6\math@{\global\math@false\ifx 7#1\ifx 8#2%
 \ifx 9#3\ifx 0#4\ifx !#5\xdef\meaning@{#6}\global\math@true\fi\fi\fi\fi\fi}}}
\newif\ifmathch@
{\uccode`7=`c \uccode`8=`h \uccode`9=`\"
 \uppercase{\gdef\mathch@#1#2#3#4#5#6\mathch@{\global\mathch@false
  \ifx 7#1\ifx 8#2\ifx 9#5\global\mathch@true\xdef\meaning@{9#6}\fi\fi\fi}}}
\newcount\classnum@
\def\getmathch@#1.#2\getmathch@{\classnum@#1 \divide\classnum@4096
 \ifcase\number\classnum@\or\or\gdef\thedots@{\dotsb@}\or
 \gdef\thedots@{\dotsb@}\fi}
\newif\ifmathbin@
{\uccode`4=`b \uccode`5=`i \uccode`6=`n
 \uppercase{\gdef\mathbin@#1#2#3{\relaxnext@
  \DNii@##1\mathbin@{\ifx\space@\next\global\mathbin@true\fi}%
 \global\mathbin@false\DN@##1\mathbin@{}%
 \ifx 4#1\ifx 5#2\ifx 6#3\DN@{\FN@\nextii@}\fi\fi\fi\next@}}}
\newif\ifmathrel@
{\uccode`4=`r \uccode`5=`e \uccode`6=`l
 \uppercase{\gdef\mathrel@#1#2#3{\relaxnext@
  \DNii@##1\mathrel@{\ifx\space@\next\global\mathrel@true\fi}%
 \global\mathrel@false\DN@##1\mathrel@{}%
 \ifx 4#1\ifx 5#2\ifx 6#3\DN@{\FN@\nextii@}\fi\fi\fi\next@}}}
\newif\ifmacro@
{\uccode`5=`m \uccode`6=`a \uccode`7=`c
 \uppercase{\gdef\macro@#1#2#3#4\macro@{\global\macro@false
  \ifx 5#1\ifx 6#2\ifx 7#3\global\macro@true
  \xdef\meaning@{\macro@@#4\macro@@}\fi\fi\fi}}}
\def\macro@@#1->#2\macro@@{#2}
\newif\ifDOTS@
\newcount\DOTSCASE@
{\uccode`6=`\\ \uccode`7=`D \uccode`8=`O \uccode`9=`T \uccode`0=`S
 \uppercase{\gdef\DOTS@#1#2#3#4#5{\global\DOTS@false\DN@##1\DOTS@{}%
  \ifx 6#1\ifx 7#2\ifx 8#3\ifx 9#4\ifx 0#5\let\next@\DOTS@@\fi\fi\fi\fi\fi
  \next@}}}
{\uccode`3=`B \uccode`4=`I \uccode`5=`X
 \uppercase{\gdef\DOTS@@#1{\relaxnext@
  \DNii@##1\DOTS@{\ifx\space@\next\global\DOTS@true\fi}%
  \DN@{\FN@\nextii@}%
  \ifx 3#1\global\DOTSCASE@\z@\else
  \ifx 4#1\global\DOTSCASE@\@ne\else
  \ifx 5#1\global\DOTSCASE@\tw@\else\DN@##1\DOTS@{}%
  \fi\fi\fi\next@}}}
\newif\ifnot@
{\uccode`5=`\\ \uccode`6=`n \uccode`7=`o \uccode`8=`t
 \uppercase{\gdef\not@#1#2#3#4{\relaxnext@
  \DNii@##1\not@{\ifx\space@\next\global\not@true\fi}%
 \global\not@false\DN@##1\not@{}%
 \ifx 5#1\ifx 6#2\ifx 7#3\ifx 8#4\DN@{\FN@\nextii@}\fi\fi\fi
 \fi\next@}}}
\newif\ifkeybin@
\def\keybin@{\keybin@true
 \ifx\next+\else\ifx\next=\else\ifx\next<\else\ifx\next>\else\ifx\next-\else
 \ifx\next*\else\ifx\next:\else\keybin@false\fi\fi\fi\fi\fi\fi\fi}
\def\dots{\RIfM@\expandafter\mdots@\else\expandafter\tdots@\fi}
\def\tdots@{\unskip\relaxnext@
 \DN@{$\m@th\mathinner{\ldotp\ldotp\ldotp}\,
   \ifx\next,\,$\else\ifx\next.\,$\else\ifx\next;\,$\else\ifx\next:\,$\else
   \ifx\next?\,$\else\ifx\next!\,$\else$ \fi\fi\fi\fi\fi\fi}%
 \ \FN@\next@}
\def\mdots@{\FN@\mdots@@}
\def\mdots@@{\gdef\thedots@{\dotso@}
 \ifx\next\boldkey\gdef\thedots@\boldkey{\boldkeydots@}\else                
 \ifx\next\boldsymbol\gdef\thedots@\boldsymbol{\boldsymboldots@}\else       
 \ifx,\next\gdef\thedots@{\dotsc}
 \else\ifx\not\next\gdef\thedots@{\dotsb@}
 \else\keybin@
 \ifkeybin@\gdef\thedots@{\dotsb@}
 \else\xdef\meaning@{\meaning\next..........}\xdef\meaning@@{\meaning@}
  \expandafter\math@\meaning@\math@
  \ifmath@
   \expandafter\mathch@\meaning@\mathch@
   \ifmathch@\expandafter\getmathch@\meaning@\getmathch@\fi                 
  \else\expandafter\macro@\meaning@@\macro@                                 
  \ifmacro@                                                                
   \expandafter\not@\meaning@\not@\ifnot@\gdef\thedots@{\dotsb@}
  \else\expandafter\DOTS@\meaning@\DOTS@
  \ifDOTS@
   \ifcase\number\DOTSCASE@\gdef\thedots@{\dotsb@}%
    \or\gdef\thedots@{\dotsi}\else\fi                                      
  \else\expandafter\math@\meaning@\math@                                   
  \ifmath@\expandafter\mathbin@\meaning@\mathbin@
  \ifmathbin@\gdef\thedots@{\dotsb@}
  \else\expandafter\mathrel@\meaning@\mathrel@
  \ifmathrel@\gdef\thedots@{\dotsb@}
  \fi\fi\fi\fi\fi\fi\fi\fi\fi\fi\fi\fi
 \thedots@}
\def\plainldots@{\mathinner{\ldotp\ldotp\ldotp}}
\def\plaincdots@{\mathinner{\cdotp\cdotp\cdotp}}
\def\dotsi{\!\plaincdots@}
\let\dotsb@\plaincdots@
\newif\ifextra@
\newif\ifrightdelim@
\def\rightdelim@{\global\rightdelim@true                                    
 \ifx\next)\else                                                            
 \ifx\next]\else
 \ifx\next\rbrack\else
 \ifx\next\}\else
 \ifx\next\rbrace\else
 \ifx\next\rangle\else
 \ifx\next\rceil\else
 \ifx\next\rfloor\else
 \ifx\next\rgroup\else
 \ifx\next\rmoustache\else
 \ifx\next\right\else
 \ifx\next\bigr\else
 \ifx\next\biggr\else
 \ifx\next\Bigr\else                                                        
 \ifx\next\Biggr\else\global\rightdelim@false
 \fi\fi\fi\fi\fi\fi\fi\fi\fi\fi\fi\fi\fi\fi\fi}
\def\extra@{%
 \global\extra@false\rightdelim@\ifrightdelim@\global\extra@true            
 \else\ifx\next$\global\extra@true                                          
 \else\xdef\meaning@{\meaning\next..........}
 \expandafter\macro@\meaning@\macro@\ifmacro@                               
 \expandafter\DOTS@\meaning@\DOTS@
 \ifDOTS@
 \ifnum\DOTSCASE@=\tw@\global\extra@true                                    
 \fi\fi\fi\fi\fi}
\newif\ifbold@
\def\dotso@{\relaxnext@
 \ifbold@
  \let\next\delayed@
  \DNii@{\extra@\plainldots@\ifextra@\,\fi}%
 \else
  \DNii@{\DN@{\extra@\plainldots@\ifextra@\,\fi}\FN@\next@}%
 \fi
 \nextii@}
\def\extrap@#1{%
 \ifx\next,\DN@{#1\,}\else
 \ifx\next;\DN@{#1\,}\else
 \ifx\next.\DN@{#1\,}\else\extra@
 \ifextra@\DN@{#1\,}\else
 \let\next@#1\fi\fi\fi\fi\next@}
\def\ldots{\DN@{\extrap@\plainldots@}%
 \FN@\next@}
\def\cdots{\DN@{\extrap@\plaincdots@}%
 \FN@\next@}

\def\dotsc{\relaxnext@
 \DN@{\ifx\next;\plainldots@\,\else
  \ifx\next.\plainldots@\,\else\extra@\plainldots@
  \ifextra@\,\fi\fi\fi}%
 \FN@\next@}
\def\cdot{\mathchar"2201 }

\message{special superscripts,}
\def\dddot#1{{\mathop{#1}\limits^{\vbox to-1.4\ex@{\kern-\tw@\ex@
 \hbox{\rm...}\vss}}}}
\def\ddddot#1{{\mathop{#1}\limits^{\vbox to-1.4\ex@{\kern-\tw@\ex@
 \hbox{\rm....}\vss}}}}
\def\sphat{^{\mathchoice{}{}%
 {\,\,\botsmash{\hbox{\lower4\ex@\hbox{$\m@th\widehat{\null}$}}}}%
 {\,\botsmash{\hbox{\lower3\ex@\hbox{$\m@th\hat{\null}$}}}}}}

\def\spacute{^{\!\botsmash{\hbox{\lower\@ne ex\hbox{\'{}}}}}}
\def\spgrave{^{\mathchoice{}{}{}{\!}%
 \botsmash{\hbox{\lower\@ne ex\hbox{\`{}}}}}}
\def\spdot{^{\hbox{\raise\ex@\hbox{\rm.}}}}
\def\spddot{^{\hbox{\raise\ex@\hbox{\rm..}}}}
\def\spdddot{^{\hbox{\raise\ex@\hbox{\rm...}}}}
\def\spddddot{^{\hbox{\raise\ex@\hbox{\rm....}}}}
\def\spbreve{^{\!\botsmash{\hbox{\lower4\ex@\hbox{\u{}}}}}}

\message{\string\text,}
\def\textonlyfont@#1#2{\def#1{\RIfM@
 \Err@{Use \string#1\space only in text}\else#2\fi}}
\textonlyfont@\rm\tenrm
\textonlyfont@\it\tenit
\textonlyfont@\sl\tensl
\textonlyfont@\bf\tenbf
\def\oldnos#1{\RIfM@{\mathcode`\,="013B \fam\@ne#1}\else
 \leavevmode\hbox{$\m@th\mathcode`\,="013B \fam\@ne#1$}\fi}
\def\text{\RIfM@\expandafter\text@\else\expandafter\text@@\fi}
\def\text@@#1{\leavevmode\hbox{#1}}
\def\mathhexbox@#1#2#3{\text{$\m@th\mathchar"#1#2#3$}}
\def\dag{{\mathhexbox@279}}
\def\ddag{{\mathhexbox@27A}}
\def\S{{\mathhexbox@278}}
\def\P{{\mathhexbox@27B}}
\newif\iffirstchoice@
\firstchoice@true
\def\text@#1{\mathchoice
 {\hbox{\everymath{\displaystyle}\def\textfonti{\the\textfont\@ne}%
  \def\textfontii{\the\textfont\tw@}\textdef@@ T#1}}
 {\hbox{\firstchoice@false
  \everymath{\textstyle}\def\textfonti{\the\textfont\@ne}%
  \def\textfontii{\the\textfont\tw@}\textdef@@ T#1}}
 {\hbox{\firstchoice@false
  \everymath{\scriptstyle}\def\textfonti{\the\scriptfont\@ne}%
  \def\textfontii{\the\scriptfont\tw@}\textdef@@ S\rm#1}}
 {\hbox{\firstchoice@false
  \everymath{\scriptscriptstyle}\def\textfonti
  {\the\scriptscriptfont\@ne}%
  \def\textfontii{\the\scriptscriptfont\tw@}\textdef@@ s\rm#1}}}
\def\textdef@@#1{\textdef@#1\rm\textdef@#1\bf\textdef@#1\sl\textdef@#1\it}
\def\rmfam{0}
\def\textdef@#1#2{%
 \DN@{\csname\expandafter\eat@\string#2fam\endcsname}%
 \if S#1\edef#2{\the\scriptfont\next@\relax}%
 \else\if s#1\edef#2{\the\scriptscriptfont\next@\relax}%
 \else\edef#2{\the\textfont\next@\relax}\fi\fi}
\scriptfont\itfam\tenit \scriptscriptfont\itfam\tenit
\scriptfont\slfam\tensl \scriptscriptfont\slfam\tensl
\newif\iftopfolded@
\newif\ifbotfolded@
\def\topfoldedtext{\topfolded@true\botfolded@false\foldedtext@}
\def\botfoldedtext{\botfolded@true\topfolded@false\foldedtext@}
\def\foldedtext{\topfolded@false\botfolded@false\foldedtext@}
\Invalid@\foldedwidth
\def\foldedtext@{\relaxnext@
 \DN@{\ifx\next\foldedwidth\let\next@\nextii@\else
  \DN@{\nextii@\foldedwidth{.3\hsize}}\fi\next@}%
 \DNii@\foldedwidth##1##2{\setbox\z@\vbox
  {\normalbaselines\hsize##1\relax
  \tolerance1600 \noindent\ignorespaces##2}\ifbotfolded@\boxz@\else
  \iftopfolded@\vtop{\unvbox\z@}\else\vcenter{\boxz@}\fi\fi}%
 \FN@\next@}
\message{math font commands,}
\def\bold{\RIfM@\expandafter\bold@\else
 \expandafter\nonmatherr@\expandafter\bold\fi}
\def\bold@#1{{\bold@@{#1}}}
\def\bold@@#1{\fam\bffam\relax#1}
\def\slanted{\RIfM@\expandafter\slanted@\else
 \expandafter\nonmatherr@\expandafter\slanted\fi}
\def\slanted@#1{{\slanted@@{#1}}}
\def\slanted@@#1{\fam\slfam\relax#1}
\def\roman{\RIfM@\expandafter\roman@\else
 \expandafter\nonmatherr@\expandafter\roman\fi}
\def\roman@#1{{\roman@@{#1}}}
\def\roman@@#1{\fam\rmfam\relax#1}
\def\italic{\RIfM@\expandafter\italic@\else
 \expandafter\nonmatherr@\expandafter\italic\fi}
\def\italic@#1{{\italic@@{#1}}}
\def\italic@@#1{\fam\itfam\relax#1}
\def\Cal{\RIfM@\expandafter\Cal@\else
 \expandafter\nonmatherr@\expandafter\Cal\fi}
\def\Cal@#1{{\Cal@@{#1}}}
\def\Cal@@#1{\noaccents@\fam\tw@#1}
\mathchardef\Gamma="0000
\mathchardef\Delta="0001
\mathchardef\Theta="0002
\mathchardef\Lambda="0003
\mathchardef\Xi="0004
\mathchardef\Pi="0005
\mathchardef\Sigma="0006
\mathchardef\Upsilon="0007
\mathchardef\Phi="0008
\mathchardef\Psi="0009
\mathchardef\Omega="000A
\mathchardef\varGamma="0100
\mathchardef\varDelta="0101
\mathchardef\varTheta="0102
\mathchardef\varLambda="0103
\mathchardef\varXi="0104
\mathchardef\varPi="0105
\mathchardef\varSigma="0106
\mathchardef\varUpsilon="0107
\mathchardef\varPhi="0108
\mathchardef\varPsi="0109
\mathchardef\varOmega="010A
\let\alloc@@\alloc@
\def\hexnumber@#1{\ifcase#1 0\or 1\or 2\or 3\or 4\or 5\or 6\or 7\or 8\or
 9\or A\or B\or C\or D\or E\or F\fi}
\def\loadmsam{%
 \font@\tenmsa=msam10
 \font@\sevenmsa=msam7
 \font@\fivemsa=msam5
 \alloc@@8\fam\chardef\sixt@@n\msafam
 \textfont\msafam=\tenmsa
 \scriptfont\msafam=\sevenmsa
 \scriptscriptfont\msafam=\fivemsa
 \edef\next{\hexnumber@\msafam}%
 \mathchardef\dabar@"0\next39
 \edef\dashrightarrow{\mathrel{\dabar@\dabar@\mathchar"0\next4B}}%
 \edef\dashleftarrow{\mathrel{\mathchar"0\next4C\dabar@\dabar@}}%
 \let\dasharrow\dashrightarrow
 \edef\ulcorner{\delimiter"4\next70\next70 }%
 \edef\urcorner{\delimiter"5\next71\next71 }%
 \edef\llcorner{\delimiter"4\next78\next78 }%
 \edef\lrcorner{\delimiter"5\next79\next79 }%
 \edef\yen{{\noexpand\mathhexbox@\next55}}%
 \edef\checkmark{{\noexpand\mathhexbox@\next58}}%
 \edef\circledR{{\noexpand\mathhexbox@\next72}}%
 \edef\maltese{{\noexpand\mathhexbox@\next7A}}%
 \global\let\loadmsam\empty}%
\def\loadmsbm{%
 \font@\tenmsb=msbm10 \font@\sevenmsb=msbm7 \font@\fivemsb=msbm5
 \alloc@@8\fam\chardef\sixt@@n\msbfam
 \textfont\msbfam=\tenmsb
 \scriptfont\msbfam=\sevenmsb \scriptscriptfont\msbfam=\fivemsb
 \global\let\loadmsbm\empty
 }
\def\widehat#1{\ifx\undefined\msbfam \DN@{362}%
  \else \setboxz@h{$\m@th#1$}%
    \edef\next@{\ifdim\wdz@>\tw@ em%
        \hexnumber@\msbfam 5B%
      \else 362\fi}\fi
  \mathaccent"0\next@{#1}}
\def\widetilde#1{\ifx\undefined\msbfam \DN@{365}%
  \else \setboxz@h{$\m@th#1$}%
    \edef\next@{\ifdim\wdz@>\tw@ em%
        \hexnumber@\msbfam 5D%
      \else 365\fi}\fi
  \mathaccent"0\next@{#1}}
\message{\string\newsymbol,}
\def\newsymbol#1#2#3#4#5{\define#1{}%
  \count@#2\relax \advance\count@\m@ne 
 \ifcase\count@
   \ifx\undefined\msafam\loadmsam\fi \let\next@\msafam
 \or \ifx\undefined\msbfam\loadmsbm\fi \let\next@\msbfam
 \else  \Err@{\Invalid@@\string\newsymbol}\let\next@\tw@\fi
 \mathchardef#1="#3\hexnumber@\next@#4#5\space}

\def\Bbb{\RIfM@\expandafter\Bbb@\else
 \expandafter\nonmatherr@\expandafter\Bbb\fi}
\def\Bbb@#1{{\Bbb@@{#1}}}
\def\Bbb@@#1{\noaccents@\fam\msbfam\relax#1}
\message{bold Greek and bold symbols,}
\def\loadbold{%
 \font@\tencmmib=cmmib10 \font@\sevencmmib=cmmib7 \font@\fivecmmib=cmmib5
 \skewchar\tencmmib'177 \skewchar\sevencmmib'177 \skewchar\fivecmmib'177
 \alloc@@8\fam\chardef\sixt@@n\cmmibfam
 \textfont\cmmibfam\tencmmib
 \scriptfont\cmmibfam\sevencmmib \scriptscriptfont\cmmibfam\fivecmmib
 \font@\tencmbsy=cmbsy10 \font@\sevencmbsy=cmbsy7 \font@\fivecmbsy=cmbsy5
 \skewchar\tencmbsy'60 \skewchar\sevencmbsy'60 \skewchar\fivecmbsy'60
 \alloc@@8\fam\chardef\sixt@@n\cmbsyfam
 \textfont\cmbsyfam\tencmbsy
 \scriptfont\cmbsyfam\sevencmbsy \scriptscriptfont\cmbsyfam\fivecmbsy
 \let\loadbold\empty
}
\def\boldnotloaded#1{\Err@{\ifcase#1\or First\else Second\fi
       bold symbol font not loaded}}
\def\mathchari@#1#2#3{\ifx\undefined\cmmibfam
    \boldnotloaded@\@ne
  \else\mathchar"#1\hexnumber@\cmmibfam#2#3\space \fi}
\def\mathcharii@#1#2#3{\ifx\undefined\cmbsyfam
    \boldnotloaded\tw@
  \else \mathchar"#1\hexnumber@\cmbsyfam#2#3\space\fi}
\edef\bffam@{\hexnumber@\bffam}
\def\boldkey#1{\ifcat\noexpand#1A%
  \ifx\undefined\cmmibfam \boldnotloaded\@ne
  \else {\fam\cmmibfam#1}\fi
 \else
 \ifx#1!\mathchar"5\bffam@21 \else
 \ifx#1(\mathchar"4\bffam@28 \else\ifx#1)\mathchar"5\bffam@29 \else
 \ifx#1+\mathchar"2\bffam@2B \else\ifx#1:\mathchar"3\bffam@3A \else
 \ifx#1;\mathchar"6\bffam@3B \else\ifx#1=\mathchar"3\bffam@3D \else
 \ifx#1?\mathchar"5\bffam@3F \else\ifx#1[\mathchar"4\bffam@5B \else
 \ifx#1]\mathchar"5\bffam@5D \else
 \ifx#1,\mathchari@63B \else
 \ifx#1-\mathcharii@200 \else
 \ifx#1.\mathchari@03A \else
 \ifx#1/\mathchari@03D \else
 \ifx#1<\mathchari@33C \else
 \ifx#1>\mathchari@33E \else
 \ifx#1*\mathcharii@203 \else
 \ifx#1|\mathcharii@06A \else
 \ifx#10\bold0\else\ifx#11\bold1\else\ifx#12\bold2\else\ifx#13\bold3\else
 \ifx#14\bold4\else\ifx#15\bold5\else\ifx#16\bold6\else\ifx#17\bold7\else
 \ifx#18\bold8\else\ifx#19\bold9\else
  \Err@{\string\boldkey\space can't be used with #1}%
 \fi\fi\fi\fi\fi\fi\fi\fi\fi\fi\fi\fi\fi\fi\fi
 \fi\fi\fi\fi\fi\fi\fi\fi\fi\fi\fi\fi\fi\fi}
\def\boldsymbol#1{%
 \DN@{\Err@{You can't use \string\boldsymbol\space with \string#1}#1}%
 \ifcat\noexpand#1A%
   \let\next@\relax
   \ifx\undefined\cmmibfam \boldnotloaded\@ne
   \else {\fam\cmmibfam#1}\fi
 \else
  \xdef\meaning@{\meaning#1.........}%
  \expandafter\math@\meaning@\math@
  \ifmath@
   \expandafter\mathch@\meaning@\mathch@
   \ifmathch@
    \expandafter\boldsymbol@@\meaning@\boldsymbol@@
   \fi
  \else
   \expandafter\macro@\meaning@\macro@
   \expandafter\delim@\meaning@\delim@
   \ifdelim@
    \expandafter\delim@@\meaning@\delim@@
   \else
    \boldsymbol@{#1}%
   \fi
  \fi
 \fi
 \next@}
\def\mathhexboxii@#1#2{\ifx\undefined\cmbsyfam
    \boldnotloaded\tw@
  \else \mathhexbox@{\hexnumber@\cmbsyfam}{#1}{#2}\fi}
\def\boldsymbol@#1{\let\next@\relax\let\next#1%
 \ifx\next\cdot\mathcharii@201 \else
 \ifx\next\prime{{\null\mathcharii@030 \null}}\else
 \ifx\next\lbrack\mathchar"4\bffam@5B \else
 \ifx\next\rbrack\mathchar"5\bffam@5D \else
 \ifx\next\{\mathcharii@466 \else
 \ifx\next\lbrace\mathcharii@466 \else
 \ifx\next\}\mathcharii@567 \else
 \ifx\next\rbrace\mathcharii@567 \else
 \ifx\next\surd{{\mathcharii@170}}\else
 \ifx\next\S{{\mathhexboxii@78}}\else
 \ifx\next\P{{\mathhexboxii@7B}}\else
 \ifx\next\dag{{\mathhexboxii@79}}\else
 \ifx\next\ddag{{\mathhexboxii@7A}}\else
 \DN@{\Err@{You can't use \string\boldsymbol\space with \string#1}#1}%
 \fi\fi\fi\fi\fi\fi\fi\fi\fi\fi\fi\fi\fi}
\def\boldsymbol@@#1.#2\boldsymbol@@{\classnum@#1 \count@@@\classnum@        
 \divide\classnum@4096 \count@\classnum@                                    
 \multiply\count@4096 \advance\count@@@-\count@ \count@@\count@@@           
 \divide\count@@@\@cclvi \count@\count@@                                    
 \multiply\count@@@\@cclvi \advance\count@@-\count@@@                       
 \divide\count@@@\@cclvi                                                    
 \multiply\classnum@4096 \advance\classnum@\count@@                         
 \ifnum\count@@@=\z@                                                        
  \count@"\bffam@ \multiply\count@\@cclvi
  \advance\classnum@\count@
  \DN@{\mathchar\number\classnum@}%
 \else
  \ifnum\count@@@=\@ne                                                      
   \ifx\undefined\cmmibfam \DN@{\boldnotloaded\@ne}%
   \else \count@\cmmibfam \multiply\count@\@cclvi
     \advance\classnum@\count@
     \DN@{\mathchar\number\classnum@}\fi
  \else
   \ifnum\count@@@=\tw@                                                    
     \ifx\undefined\cmbsyfam
       \DN@{\boldnotloaded\tw@}%
     \else
       \count@\cmbsyfam \multiply\count@\@cclvi
       \advance\classnum@\count@
       \DN@{\mathchar\number\classnum@}%
     \fi
  \fi
 \fi
\fi}
\newif\ifdelim@
\newcount\delimcount@
{\uccode`6=`\\ \uccode`7=`d \uccode`8=`e \uccode`9=`l
 \uppercase{\gdef\delim@#1#2#3#4#5\delim@
  {\delim@false\ifx 6#1\ifx 7#2\ifx 8#3\ifx 9#4\delim@true
   \xdef\meaning@{#5}\fi\fi\fi\fi}}}
\def\delim@@#1"#2#3#4#5#6\delim@@{\if#32%
\let\next@\relax
 \ifx\undefined\cmbsyfam \boldnotloaded\@ne
 \else \mathcharii@#2#4#5\space \fi\fi}
\def\vert{\delimiter"026A30C }
\def\Vert{\delimiter"026B30D }
\let\|\Vert

\def\boldkeydots@#1{\bold@true\let\next=#1\let\delayed@=#1\mdots@@
 \boldkey#1\bold@false}  
\def\boldsymboldots@#1{\bold@true\let\next#1\let\delayed@#1\mdots@@
 \boldsymbol#1\bold@false}
\message{Euler fonts,}

\def\frak{\mathfont@\frak}

\def\loadmathfont#1{%
   \expandafter\font@\csname ten#1\endcsname=#110
   \expandafter\font@\csname seven#1\endcsname=#17
   \expandafter\font@\csname five#1\endcsname=#15
   \edef\next{\noexpand\alloc@@8\fam\chardef\sixt@@n
     \expandafter\noexpand\csname#1fam\endcsname}%
   \next
   \textfont\csname#1fam\endcsname \csname ten#1\endcsname
   \scriptfont\csname#1fam\endcsname \csname seven#1\endcsname
   \scriptscriptfont\csname#1fam\endcsname \csname five#1\endcsname
   \expandafter\def\csname #1\expandafter\endcsname\expandafter{%
      \expandafter\mathfont@\csname#1\endcsname}%
 \expandafter\gdef\csname load#1\endcsname{}%
}
\def\mathfont@#1{\RIfM@\expandafter\mathfont@@\expandafter#1\else
  \expandafter\nonmatherr@\expandafter#1\fi}
\def\mathfont@@#1#2{{\mathfont@@@#1{#2}}}
\def\mathfont@@@#1#2{\noaccents@
   \fam\csname\expandafter\eat@\string#1fam\endcsname
   \relax#2}
\message{math accents,}
\def\accentclass@{7}
\def\noaccents@{\def\accentclass@{0}}
\def\makeacc@#1#2{\def#1{\mathaccent"\accentclass@#2 }}
\makeacc@\hat{05E}
\makeacc@\check{014}
\makeacc@\tilde{07E}
\makeacc@\acute{013}
\makeacc@\grave{012}
\makeacc@\dot{05F}
\makeacc@\ddot{07F}
\makeacc@\breve{015}
\makeacc@\bar{016}

\newcount\skewcharcount@
\newcount\familycount@
\def\theskewchar@{\familycount@\@ne
 \global\skewcharcount@\the\skewchar\textfont\@ne                           
 \ifnum\fam>\m@ne\ifnum\fam<16
  \global\familycount@\the\fam\relax
  \global\skewcharcount@\the\skewchar\textfont\the\fam\relax\fi\fi          
 \ifnum\skewcharcount@>\m@ne
  \ifnum\skewcharcount@<128
  \multiply\familycount@256
  \global\advance\skewcharcount@\familycount@
  \global\advance\skewcharcount@28672
  \mathchar\skewcharcount@\else
  \global\skewcharcount@\m@ne\fi\else
 \global\skewcharcount@\m@ne\fi}                                            
\newcount\pointcount@
\def\getpoints@#1.#2\getpoints@{\pointcount@#1 }
\newdimen\accentdimen@
\newcount\accentmu@
\def\dimentomu@{\multiply\accentdimen@ 100
 \expandafter\getpoints@\the\accentdimen@\getpoints@
 \multiply\pointcount@18
 \divide\pointcount@\@m
 \global\accentmu@\pointcount@}
\def\Makeacc@#1#2{\def#1{\RIfM@\DN@{\mathaccent@
 {"\accentclass@#2 }}\else\DN@{\nonmatherr@{#1}}\fi\next@}}
\def\unbracefonts@{\let\Cal@\Cal@@\let\roman@\roman@@\let\bold@\bold@@
 \let\slanted@\slanted@@}
\def\mathaccent@#1#2{\ifnum\fam=\m@ne\xdef\thefam@{1}\else
 \xdef\thefam@{\the\fam}\fi                                                 
 \accentdimen@\z@                                                           
 \setboxz@h{\unbracefonts@$\m@th\fam\thefam@\relax#2$}
 \ifdim\accentdimen@=\z@\DN@{\mathaccent#1{#2}}
  \setbox@ne\hbox{\unbracefonts@$\m@th\fam\thefam@\relax#2\theskewchar@$}
  \setbox\tw@\hbox{$\m@th\ifnum\skewcharcount@=\m@ne\else
   \mathchar\skewcharcount@\fi$}
  \global\accentdimen@\wd@ne\global\advance\accentdimen@-\wdz@
  \global\advance\accentdimen@-\wd\tw@                                     
  \global\multiply\accentdimen@\tw@
  \dimentomu@\global\advance\accentmu@\@ne                                 
 \else\DN@{{\mathaccent#1{#2\mkern\accentmu@ mu}%
    \mkern-\accentmu@ mu}{}}\fi                                             
 \next@}\Makeacc@\Hat{05E}
\Makeacc@\Check{014}
\Makeacc@\Tilde{07E}
\Makeacc@\Acute{013}
\Makeacc@\Grave{012}
\Makeacc@\Dot{05F}
\Makeacc@\Ddot{07F}
\Makeacc@\Breve{015}
\Makeacc@\Bar{016}
\def\Vec{\RIfM@\DN@{\mathaccent@{"017E }}\else
 \DN@{\nonmatherr@\Vec}\fi\next@}
\def\accentedsymbol#1#2{\csname newbox\expandafter\endcsname
  \csname\expandafter\eat@\string#1@box\endcsname
 \expandafter\setbox\csname\expandafter\eat@
  \string#1@box\endcsname\hbox{$\m@th#2$}\define
  #1{\copy\csname\expandafter\eat@\string#1@box\endcsname{}}}
\message{roots,}
\def\sqrt#1{\radical"270370 {#1}}
\let\underline@\underline
\let\overline@\overline
\def\underline#1{\underline@{#1}}
\def\overline#1{\overline@{#1}}
\Invalid@\leftroot
\Invalid@\uproot
\newcount\uproot@
\newcount\leftroot@
\def\root{\relaxnext@
  \DN@{\ifx\next\uproot\let\next@\nextii@\else
   \ifx\next\leftroot\let\next@\nextiii@\else
   \let\next@\plainroot@\fi\fi\next@}%
  \DNii@\uproot##1{\uproot@##1\relax\FN@\nextiv@}%
  \def\nextiv@{\ifx\next\space@\DN@. {\FN@\nextv@}\else
   \DN@.{\FN@\nextv@}\fi\next@.}%
  \def\nextv@{\ifx\next\leftroot\let\next@\nextvi@\else
   \let\next@\plainroot@\fi\next@}%
  \def\nextvi@\leftroot##1{\leftroot@##1\relax\plainroot@}%
   \def\nextiii@\leftroot##1{\leftroot@##1\relax\FN@\nextvii@}%
  \def\nextvii@{\ifx\next\space@
   \DN@. {\FN@\nextviii@}\else
   \DN@.{\FN@\nextviii@}\fi\next@.}%
  \def\nextviii@{\ifx\next\uproot\let\next@\nextix@\else
   \let\next@\plainroot@\fi\next@}%
  \def\nextix@\uproot##1{\uproot@##1\relax\plainroot@}%
  \bgroup\uproot@\z@\leftroot@\z@\FN@\next@}
\def\plainroot@#1\of#2{\setbox\rootbox\hbox{$\m@th\scriptscriptstyle{#1}$}%
 \mathchoice{\r@@t\displaystyle{#2}}{\r@@t\textstyle{#2}}
 {\r@@t\scriptstyle{#2}}{\r@@t\scriptscriptstyle{#2}}\egroup}
\def\r@@t#1#2{\setboxz@h{$\m@th#1\sqrt{#2}$}%
 \dimen@\ht\z@\advance\dimen@-\dp\z@
 \setbox@ne\hbox{$\m@th#1\mskip\uproot@ mu$}\advance\dimen@ 1.667\wd@ne
 \mkern-\leftroot@ mu\mkern5mu\raise.6\dimen@\copy\rootbox
 \mkern-10mu\mkern\leftroot@ mu\boxz@}
\def\boxed#1{\setboxz@h{$\m@th\displaystyle{#1}$}\dimen@.4\ex@
 \advance\dimen@3\ex@\advance\dimen@\dp\z@
 \hbox{\lower\dimen@\hbox{%
 \vbox{\hrule height.4\ex@
 \hbox{\vrule width.4\ex@\hskip3\ex@\vbox{\vskip3\ex@\boxz@\vskip3\ex@}%
 \hskip3\ex@\vrule width.4\ex@}\hrule height.4\ex@}%
 }}}
\message{commutative diagrams,}
\let\ampersand@\relax
\newdimen\minaw@
\minaw@11.11128\ex@
\newdimen\minCDaw@
\minCDaw@2.5pc
\def\minCDarrowwidth#1{\RIfMIfI@\onlydmatherr@\minCDarrowwidth
 \else\minCDaw@#1\relax\fi\else\onlydmatherr@\minCDarrowwidth\fi}
\newif\ifCD@
\def\CD{\bgroup\vspace@\relax\let\ampersand@&\iffalse}\fi
 \CD@true\vcenter\bgroup\Let@\tabskip\z@skip\baselineskip20\ex@
 \lineskip3\ex@\lineskiplimit3\ex@\halign\bgroup
 &\hfill$\m@th##$\hfill\crcr}
\def\endCD{\crcr\egroup\egroup\egroup}
\newdimen\bigaw@
\atdef@>#1>#2>{\ampersand@                                                  
 \setboxz@h{$\m@th\ssize\;{#1}\;\;$}
 \setbox@ne\hbox{$\m@th\ssize\;{#2}\;\;$}
 \setbox\tw@\hbox{$\m@th#2$}
 \ifCD@\global\bigaw@\minCDaw@\else\global\bigaw@\minaw@\fi                 
 \ifdim\wdz@>\bigaw@\global\bigaw@\wdz@\fi
 \ifdim\wd@ne>\bigaw@\global\bigaw@\wd@ne\fi                                
 \ifCD@\enskip\fi                                                           
 \ifdim\wd\tw@>\z@
  \mathrel{\mathop{\hbox to\bigaw@{\rightarrowfill@\displaystyle}}%
    \limits^{#1}_{#2}}
 \else\mathrel{\mathop{\hbox to\bigaw@{\rightarrowfill@\displaystyle}}%
    \limits^{#1}}\fi                                                        
 \ifCD@\enskip\fi                                                          
 \ampersand@}                                                              
\atdef@<#1<#2<{\ampersand@\setboxz@h{$\m@th\ssize\;\;{#1}\;$}%
 \setbox@ne\hbox{$\m@th\ssize\;\;{#2}\;$}\setbox\tw@\hbox{$\m@th#2$}%
 \ifCD@\global\bigaw@\minCDaw@\else\global\bigaw@\minaw@\fi
 \ifdim\wdz@>\bigaw@\global\bigaw@\wdz@\fi
 \ifdim\wd@ne>\bigaw@\global\bigaw@\wd@ne\fi
 \ifCD@\enskip\fi
 \ifdim\wd\tw@>\z@
  \mathrel{\mathop{\hbox to\bigaw@{\leftarrowfill@\displaystyle}}%
       \limits^{#1}_{#2}}\else
  \mathrel{\mathop{\hbox to\bigaw@{\leftarrowfill@\displaystyle}}%
       \limits^{#1}}\fi
 \ifCD@\enskip\fi\ampersand@}
\begingroup
 \catcode`\~=\active \lccode`\~=`\@
 \lowercase{%
  \global\atdef@)#1)#2){~>#1>#2>}
  \global\atdef@(#1(#2({~<#1<#2<}}
\endgroup
\atdef@ A#1A#2A{\llap{$\m@th\vcenter{\hbox
 {$\ssize#1$}}$}\Big\uparrow\rlap{$\m@th\vcenter{\hbox{$\ssize#2$}}$}&&}
\atdef@ V#1V#2V{\llap{$\m@th\vcenter{\hbox
 {$\ssize#1$}}$}\Big\downarrow\rlap{$\m@th\vcenter{\hbox{$\ssize#2$}}$}&&}
\atdef@={&\enskip\mathrel
 {\vbox{\hrule width\minCDaw@\vskip3\ex@\hrule width
 \minCDaw@}}\enskip&}
\atdef@|{\Big\Vert&&}
\atdef@\vert{\Big\Vert&&}
\def\pretend#1\haswidth#2{\setboxz@h{$\m@th\scriptstyle{#2}$}\hbox
 to\wdz@{\hfill$\m@th\scriptstyle{#1}$\hfill}}
\message{poor man's bold,}
\def\pmb{\RIfM@\expandafter\mathpalette\expandafter\pmb@\else
 \expandafter\pmb@@\fi}
\def\pmb@@#1{\leavevmode\setboxz@h{#1}%
   \dimen@-\wdz@
   \kern-.5\ex@\copy\z@
   \kern\dimen@\kern.25\ex@\raise.4\ex@\copy\z@
   \kern\dimen@\kern.25\ex@\box\z@
}
\def\binrel@@#1{\ifdim\wd2<\z@\mathbin{#1}\else\ifdim\wd\tw@>\z@
 \mathrel{#1}\else{#1}\fi\fi}
\newdimen\pmbraise@
\def\pmb@#1#2{\setbox\thr@@\hbox{$\m@th#1{#2}$}%
 \setbox4\hbox{$\m@th#1\mkern.5mu$}\pmbraise@\wd4\relax
 \binrel@{#2}%
 \dimen@-\wd\thr@@
   \binrel@@{%
   \mkern-.8mu\copy\thr@@
   \kern\dimen@\mkern.4mu\raise\pmbraise@\copy\thr@@
   \kern\dimen@\mkern.4mu\box\thr@@
}}
\def\documentstyle#1{\W@{}\input #1.sty\relax}
\message{syntax check,}
\font\dummyft@=dummy
\fontdimen1 \dummyft@=\z@
\fontdimen2 \dummyft@=\z@
\fontdimen3 \dummyft@=\z@
\fontdimen4 \dummyft@=\z@
\fontdimen5 \dummyft@=\z@
\fontdimen6 \dummyft@=\z@
\fontdimen7 \dummyft@=\z@
\fontdimen8 \dummyft@=\z@
\fontdimen9 \dummyft@=\z@
\fontdimen10 \dummyft@=\z@
\fontdimen11 \dummyft@=\z@
\fontdimen12 \dummyft@=\z@
\fontdimen13 \dummyft@=\z@
\fontdimen14 \dummyft@=\z@
\fontdimen15 \dummyft@=\z@
\fontdimen16 \dummyft@=\z@
\fontdimen17 \dummyft@=\z@
\fontdimen18 \dummyft@=\z@
\fontdimen19 \dummyft@=\z@
\fontdimen20 \dummyft@=\z@
\fontdimen21 \dummyft@=\z@
\fontdimen22 \dummyft@=\z@
\def\fontlist@{\\{\tenrm}\\{\sevenrm}\\{\fiverm}\\{\teni}\\{\seveni}%
 \\{\fivei}\\{\tensy}\\{\sevensy}\\{\fivesy}\\{\tenex}\\{\tenbf}\\{\sevenbf}%
 \\{\fivebf}\\{\tensl}\\{\tenit}}
\def\font@#1=#2 {\rightappend@#1\to\fontlist@\font#1=#2 }
\def\dodummy@{{\def\\##1{\global\let##1\dummyft@}\fontlist@}}
\def\nopages@{\output{\setbox\z@\box\@cclv \deadcycles\z@}%
 \alloc@5\toks\toksdef\@cclvi\output}
\let\galleys\nopages@
\newif\ifsyntax@
\newcount\countxviii@
\def\syntax{\syntax@true\dodummy@\countxviii@\count18
 \loop\ifnum\countxviii@>\m@ne\textfont\countxviii@=\dummyft@
 \scriptfont\countxviii@=\dummyft@\scriptscriptfont\countxviii@=\dummyft@
 \advance\countxviii@\m@ne\repeat                                           
 \dummyft@\tracinglostchars\z@\nopages@\frenchspacing\hbadness\@M}
\def\first@#1#2\end{#1}
\def\printoptions{\W@{Do you want S(yntax check),
  G(alleys) or P(ages)?}%
 \message{Type S, G or P, followed by <return>: }%
 \begingroup 
 \endlinechar\m@ne 
 \read\m@ne to\ans@
 \edef\ans@{\uppercase{\def\noexpand\ans@{%
   \expandafter\first@\ans@ P\end}}}%
 \expandafter\endgroup\ans@
 \if\ans@ P
 \else \if\ans@ S\syntax
 \else \if\ans@ G\galleys
 \else\message{? Unknown option: \ans@; using the `pages' option.}%
 \fi\fi\fi}
\def\alloc@#1#2#3#4#5{\global\advance\count1#1by\@ne
 \ch@ck#1#4#2\allocationnumber=\count1#1
 \global#3#5=\allocationnumber
 \ifalloc@\wlog{\string#5=\string#2\the\allocationnumber}\fi}
\def\document{\def\alloclist@{}\def\fontlist@{}}

\let\footnote\undefined
\let\=\undefined
\let\>\undefined

\catcode`\@=\active
\message{... finished}

\nopagenumbers
%
\topinsert
\vskip 3.2 truecm
\endinsert
\centerline{\bigbfont SPIN-PARTICLE CONNECTIONS}
\vskip 20 truept
\centerline{\bigifont C.D. Batista and G. Ortiz }
\vskip 8 truept
\centerline{\bigrfont Theoretical Division, 
Los Alamos National Laboratory} 
\vskip 2 truept
\centerline{\bigrfont Los Alamos, NM 87545, USA} 
\vskip 1.8 truecm

\centerline{\bf 1.  INTRODUCTION}
\vskip 12 truept

Group theory and geometry have been fundamental to the physics of the
Twentieth Century. The notion of symmetry has shaped our current
conception of nature; however, nature is also full of symmetry
breakings. Therefore, understanding the idea of invariance and its
corresponding conservation laws is as fundamental as determining the
causes that prevent such harmony and leads to more complex behavior. 
These concepts are at the heart of the field of quantum phase
transitions that studies the changes that can occur in the macroscopic
properties of matter at zero temperature due to changes in the
parameters characterizing the system.

On the other hand, the notion of algebra and its homomorphisms have
also been essential to unravel hidden structures in theoretical
physics: Internal symmetries which are hidden in one algebraic
representation of a model become manifest in the other one. In 1928
Jordan and Wigner [1] made a first step relating quantum spin
$S=\frac{1}{2}$ degrees of freedom to particles with fermion
statistics, an application being the mapping of the isotropic XY model
describing planar magnets onto a tight-binding spinless fermion model
which can be exactly solved in one spatial dimension. From the group
theoretical viewpoint, an internal $U(1)$ continuous symmetry (related
to particle number conservation), for instance, is evidenced in the
fermion representation of the XY model which was hidden in the spin
representation. Overall, what Jordan and Wigner established was an
isomorphism of $*$-algebras, i.e., an isomorphism between the Pauli and
fermion algebras.

The three basis elements $S_j^\mu (\mu=x,y,z)$ (linear and Hermitian
operators) of the Lie algebra $su(2)$ for each lattice site $j$ ($j
=1,\cdots,L$) satisfy the equal-time commutation relations [2]
$$
\left [ S_j^\mu, S_k^\nu \right ] = i \delta_{j k} \epsilon_{\mu \nu
\lambda} S_j^\lambda \ , \eqno(1)
$$
with $\epsilon$ the totally antisymmetric Levi-Civita symbol.
Equivalently, in terms of the ladder operators $S_j^\pm = S_j^x \pm i
S_j^y$
$$
\left [ S_j^+, S_j^- \right ] = 2 S_j^z \ ,  \  \left [ S_j^z, S_j^\pm
\right ] = \pm S_j^\pm \ ,  \ \left \{ S_j^+, S_j^- \right \} = 2
\left( S(S+1)- (S_j^z)^2 \right ) \ , \eqno(2)
$$
where $2S+1$ is the dimension of the irreducible representation $S$.
Mathematically, the Jordan-Wigner transformation [1] involves the
$S$=$\frac{1}{2}$ irreducible representation of the Lie group $SU(2)$.
The mapping itself establishes the isomorphism of algebras: $S_j^\mu =
{\Cal F}(\{c_m^{\;}\},\{c_m^\dagger\})$, where ${\Cal F}$ is an
operator function of the ``spinless'' fermions
$(c_m^{\;},c_m^\dagger)$, for different modes $m$ ($m =1,\cdots,L$),
satisfying canonical anticommutation relations
$$
\{ c^{\;}_{j}, c^{\;}_{m} \} =
\{ c^\dagger_{j}, c^\dagger_{m} \} = 0  \ , \ 
\{ c^{\;}_{j}, c^\dagger_{m}  \} = \delta_{jm} \ . \eqno(3)
$$

Up until now, no one was able to generalize Jordan and Wigner's
findings for arbitrary spin $S$, spatial dimension and particle
statistics. In the present manuscript we generalize their spin-fermion
mapping to any irreducible representation $S$. Our mappings are valid
for regular lattices in any spatial dimension $d$ and particle
statistics. The significance of these transformations is that they help
us understand various aspects of the same physical system by
transforming intricate interaction terms in one representation into
simpler ones in the other. Problems which seem untractable can even be
exactly solved after the mapping. In other cases, new and better
approximations can, in principle, be realized since fundamental
symmetries which are hidden in one representation are manifest in the
other. 

From a physical viewpoint what our spin-particle transformations
achieve is an exact connection between models of localized quantum
spins $S$ to models of itinerant particles with ($2S=N_f$) color
degrees of freedom or ``effective'' spin $s=S-\frac{1}{2}$. As we will
see this dictionary turns out to be extremely useful to understanding
the various complex quantum phases that arise in models of itinerant
strongly correlated systems,  as a result of the competing
interactions. In the next Section, we start by analyzing the
one-dimensional $S$=1 case. Then, we will show a generalization to
arbitrary spin, spatial dimension and, finally, particle statistics
(i.e., general spin-anyon mappings for the case $d \leq 2$). We
illustrate the power of these transformations by showing exact
solutions to 1$d$ lattice models previously unsolved by standard
techniques. We also present a proof of the existence of the Haldane gap
in $S$=1 bilinear nearest-neighbors Heisenberg spin chains and discuss
the relevance of the mapping to models of strongly correlated
electrons. 

\vskip 28 truept

\centerline{\bf 2.  $S=1$ MAPPING}
\vskip 12 truept

To gain some intuition about the mapping between particles and spins,
it is instructive to start by analyzing the local Hilbert spaces
associated with each world, i.e., the Hilbert spaces associated with a
single site (mode) in each representation. In the traditional
Jordan-Wigner transformation [1], a spin $S=\frac{1}{2}$ is mapped onto
a spinless fermion. Both local Hilbert spaces are shown in Fig. 1 (a),
and from that figure it is clear that one can make a one-to-one mapping
between the states of both spaces. The convention used by Jordan and
Wigner was to map the up (down) state into the occupied (empty) state.
The next step consists of mapping the corresponding operators acting on
one site. $S_j^z$ and $n_j={c}^\dagger_{j}{c}^{\;}_{j}$ are  diagonal
operators in each representation and they are related through the
following expression
$$
S^z_j = n_{j} - \frac{1}{2} \ .\eqno(4)
$$
\topinsert
\input psfig.sty
\centerline{\hskip0mm
\psfig{figure=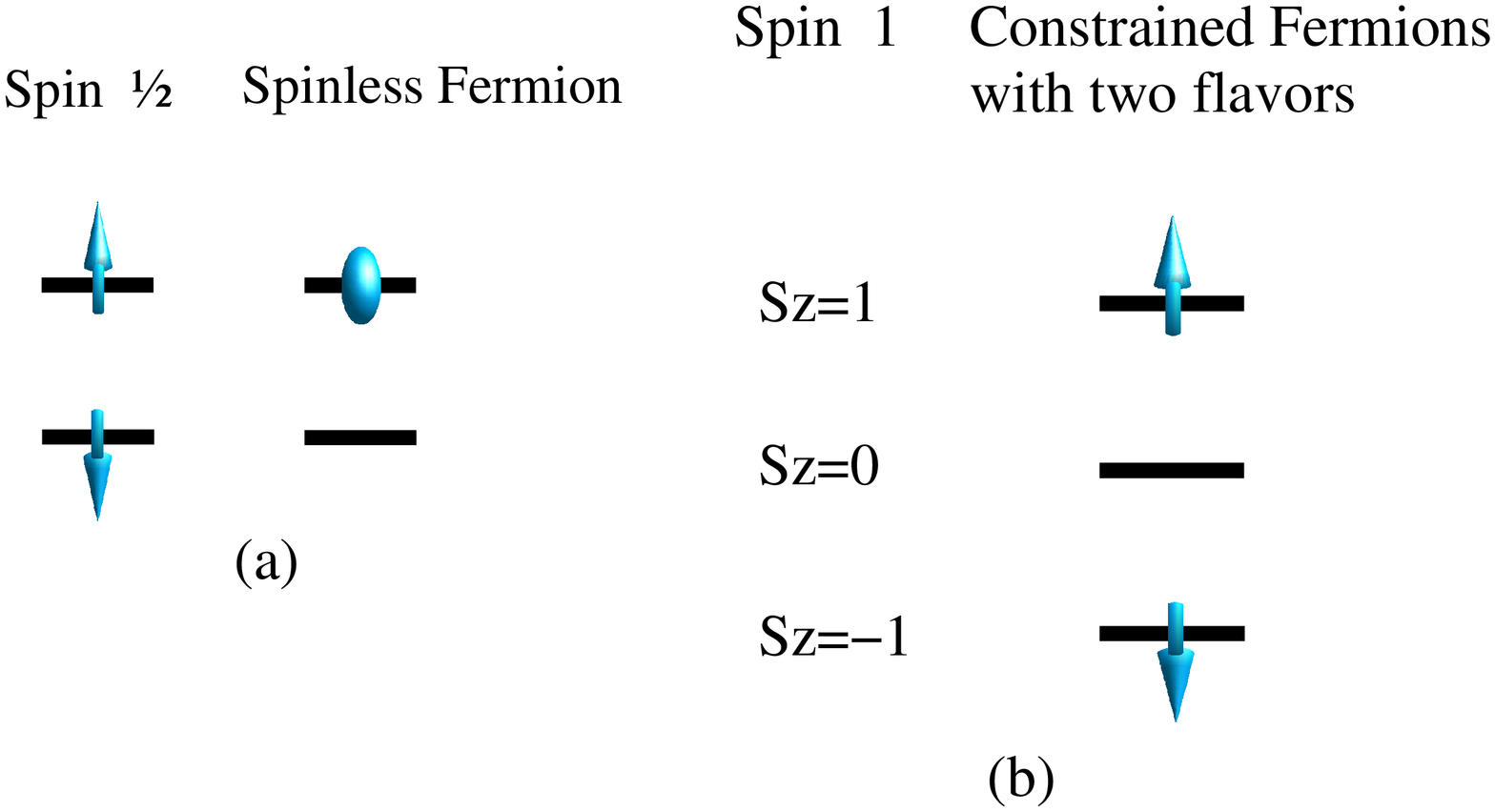,height=7truecm,width=11truecm,angle=270}}
\vskip -0.0truecm 
\noindent
{\bf Figure 1.} 
Local Hilbert spaces for the spin and fermion worlds at site (mode)
$j$: (a) $S=\frac{1}{2}$, (b) $S=1$.
\vskip 6truept
\endinsert
\noindent
$S^+_j$ and $S^-_j$ are non-diagonal in the present basis and must be
related to ${c}^\dagger_{j}$ and ${c}^{\;}_{j}$. If we consider the
problem on a lattice, the fermions can be permuted; however, the
commutation relations of the two representations are different. To take
into account the transmutation of statistics, one must introduce a
non-local operator $K_j$. Then, the transformation for the non-diagonal
operators is [1]
$$
S^+_j =  \ {c}^\dagger_{j } \ K_j  \ , \ 
S^-_j =  \ K_j^\dagger \ {c}^{\;}_{j} \ , \eqno(5)
$$
where $K_j = \exp[i \pi \sum_{k<j}{n}_k]$. This operator introduces a
negative sign each time one fermion hops over the fermion at site $j$.
This negative sign cancels the sign coming from the fermionic
statistics. The result is the transmutation of fermionic
anticommutation relations into spin commutation relations. Note that
$K_j$ does not appear in the expression for $S_j^z$ since $n_j$ is
bilinear in the fermion operators. 

This idea can be generalized to $S=1$ following similar steps to the
ones described above. Figure 1 (b) shows the two local Hilbert spaces.
In this case, the particle Hilbert space corresponds to
spin-$\frac{1}{2}$ (two flavors) fermions with the constraint of no
double occupancy. This constraint  can be taken into account by
introducing the Hubbard operators $\bar{c}^\dagger_{j \sigma}=
c^\dagger_{j \sigma} (1 - {n}_{j \bar{\sigma}})$ and $\bar{c}^{\;}_{j
\sigma}= (1 - {n}_{j \bar{\sigma}}) \, c^{\;}_{j \sigma}$
($\sigma=1,-1$; $\bar{\sigma}=-\sigma$), which form a subalgebra of the
so-called double graded algebra $Spl(1,2)$ [3]. From Fig. 1 (b) one
realizes that $S^z_j$ is the difference between the occupation numbers
of the two different fermion flavors. $S^+_j$ must be a linear
combination of annihilation and creation operators since we need to
annihilate (see Fig. 1 (b)) one fermion to go from $S^z_j=-1$ to the
$S^z_j=0$ state and to create the other fermion to go from  $S^z_j=0$
to $S^z_j=1$. To simplify notation we introduce the following composite
operators
$$
f_j^\dagger = \bar{c}^\dagger_{j 1} + \bar{c}^{\;}_{j \bar{1}} \ , \ 
f_j^{\;} = \bar{c}^{\;}_{j 1} + \bar{c}^\dagger_{j \bar{1}} \ .
\eqno(6)
$$
For spins on a lattice we again fermionize the spins and reproduce the
correct spin algebra with the following transformation
$$
S^+_j = \sqrt{2} \ (\bar{c}^\dagger_{j 1} \ K_j + K_j^\dagger \
\bar{c}^{\;}_{j \bar{1}}) \ , \ 
S^-_j = \sqrt{2} \ (K_j^\dagger \ \bar{c}^{\;}_{j 1} +
\bar{c}^\dagger_{j \bar{1}} \ K_j) \ , \ 
S^z_j = \bar{n}_{j 1} - \bar{n}_{j \bar{1}} \ , \eqno(7)
$$
whose inverse manifests the nonlocal character of the mapping
$$
\eqalignno{
f^\dagger_j &= \frac{1}{\sqrt{2}} \ \exp[i \pi \sum_{k<j} (S_k^z)^2] \ 
S_j^+  \ \ , \ \
f^{\;}_j = \frac{1}{\sqrt{2}} \exp[-i \pi \sum_{k<j} (S_k^z)^2] \ S_j^-
\ ,  \cr
\bar{c}^\dagger_{j 1}&= S_j^z \ f^\dagger_j \ \ , \ \
\bar{c}^{\;}_{j 1} =  f^{\;}_j \ S_j^z \ \ , \ \ 
\bar{c}^\dagger_{j \bar{1}}= - S_j^z \ f^{\;}_j \ \ , \ \ \bar{c}^{\;}_{j
\bar{1}} = -f^\dagger_j \ S_j^z \ ,  &(8)
}
$$
where the string operators $K_j = \exp[i \pi \sum_{k<j}\bar{n}_k] =
\prod_{k<j} \prod_\sigma (1 - 2 \bar{n}_{k \sigma})$ are the natural
generalizations of the ones introduced before [4]. The number operators
are defined as $\bar{n}_k = \bar{n}_{k 1} + \bar{n}_{k \bar{1}}$
($\bar{n}_{k \sigma} = \bar{c}^\dagger_{k \sigma} \bar{c}^{\;}_{k
\sigma}$). These $f$-operators have the remarkable property that
$$
\left \{ f_j^\dagger, f_j^{\;} \right \} =  \left \{ S_j^+, S_j^- \right
\} \ , \eqno(9)
$$
which suggests an analogy between spin operators and ``constrained''
fermions.

\vskip 28 truept

\centerline{\bf 3.  $S=1$ MODELS}
\vskip 12 truept

\centerline{\bf 3.1  Haldane Systems}
\vskip 12 truept

Generically, half-odd integer spin chains have a qualitatively
different excitation spectrum than integer spin chains. The Lieb,
Schultz, Mattis and Affleck theorem [5] establishes that the half-odd
integer antiferromagnetic (AF) bilinear nearest-neighbors (NN)
Heisenberg chain is gapless if the ground state is non-degenerate. The
same model with integer spins is conjectured to have a Haldane gap
[6]. To understand the origin of the Haldane gap we analyze the form of
the 1$d$ $S$=1 XXZ Hamiltonian using the above representation (an
overall omitted constant $J$ $>$ 0 determines the energy scale) 
$$
H_{\text {\rm xxz}} = \sum_j S^z_j S^z_{j+1} + \Delta \left ( S^x_j
S^x_{j+1}  + S^y_j S^y_{j+1} \right )  = \sum_j H_j^z + H_j^{\text {\rm
xx}} \ . \eqno(10)
$$
It is easy to show that the constrained fermion version of this
Hamiltonian is a ($S=$$\frac{1}{2}$) $t$-$J_z$ model [7] plus 
particle non-conserving terms which break the $U(1)$ symmetry
$$
H_{\text {\rm xxz}} =\sum_j (\bar{n}_{j 1} - \bar{n}_{j
\bar{1}})(\bar{n}_{j+1 1}  - \bar{n}_{j+1 \bar{1}}) + \Delta \sum_{j
\sigma} \left (  \bar{c}^\dagger_{j \sigma} \bar{c}^{\;}_{j+1 \sigma} +
\bar{c}^\dagger_{j \sigma} \bar{c}^\dagger_{j+1 \bar{\sigma}} + {\text
H.c.} \right ) \ . \eqno(11)
$$
 
The charge spectrum of the ($S=\frac{1}{2}$) $t$-$J_z$ model is gapless
but the spin spectrum is gapped due to the explicitly broken $SU(2)$
symmetry (Luther-Emery liquid) [7]. The proof of the existence of this
spin gap is given in Appendix I. Therefore, the spectrum of the $S$=1
Hamiltonian associated with the $t$-$J_z$ model, with $t=-\Delta$ and
$J_z=4$, (which has only spin excitations) is gapless. Hence the term
which explicitly breaks $U(1)$ must be responsible for the opening of
the Haldane gap. We can prove this by  considering the perturbative
effect that the interaction $\eta \sum_{j \sigma} (\bar{c}^\dagger_{j
\sigma}\bar{c}^\dagger_{j+1 \bar{\sigma}} + {\text {\rm H.c.}})$ has on
the $t$-$J_z$ Hamiltonian. To linear order in $\eta$ ($>0$), Eq. (11)
maps onto the ($S=\frac{1}{2}$) XYZ model with ${\Cal J}_x=2 (\eta +
\Delta)$, ${\Cal J}_y= -2(\eta - \Delta)$, and ${\Cal J}_z=-1$. To prove
this statement we need to explain first how the lowest energy subspace
of the $t$-$J_z$ Hamiltonian can be mapped onto a spinless $t$-$V$
model (the complete demonstration is presented in Ref. [7]).  

The $t$-$J_z$ Hamiltonian represents a hole-doped Ising model
$$
H_{t-J_z} = \hat{H}_{J_z} + \hat{T} = J_z \sum_j S^z_j S^z_{j+1}- t
\sum_{j\sigma} \left ( \bar{c}^\dagger_{j \sigma} \bar{c}^{\;}_{j+1 
\sigma} + {\text {\rm H.c.}} \right ) \ . \eqno(12)
$$
Consider the set of parent states with $M$ holes and
$L-M=N_{\uparrow}+N_{\downarrow}$ quantum particles
($\sigma=\uparrow,\downarrow$), $| \Phi_0 (N_{\uparrow
\uparrow},N_{\uparrow \downarrow}) \rangle$, defined as
$$
| \Phi_0 (N_{\uparrow \uparrow},N_{\uparrow \downarrow}) \rangle = |
\underset{L - M}\to{\underbrace{\uparrow \uparrow \downarrow \uparrow
\downarrow \cdots}} \ \underset{M}\to{\underbrace{\circ_{\;} \! \!\circ
\circ \circ \circ \cdots}} \ \rangle \ , \eqno(13)
$$
where $N_{\uparrow \uparrow}$ ($N_{\uparrow \downarrow}$) is the number
of ferro (antiferro)-magnetic links ($N_{\uparrow \uparrow}+
N_{\uparrow \downarrow}= L - M -1$). These states are eigenstates of
the magnetic part of $H_{t-J_z}$, $\hat{H}_{J_z}$, with energy
$E_M(N_{\uparrow \uparrow},N_{\uparrow \downarrow})=  J_z (N_{\uparrow
\uparrow}-N_{\uparrow \downarrow})/4$, and $z$-component of the total
spin $(N_{\uparrow}-N_{\downarrow})/2$.

From a given parent state one can generate a subspace of the Hilbert
space, ${\Cal M}(N_{\uparrow},N_{\uparrow \uparrow},N_{\uparrow
\downarrow})$, by applying the hopping operators $\hat{T}_{j,\sigma}=
c^{\dagger}_{j \sigma} c^{\;}_{j+1 \sigma} + {\text {\rm H.c.}}$
($j=1,\cdots,L-1$) to the parent state and its descendants
$$
| \Phi_1(N_{\uparrow \uparrow},N_{\uparrow \downarrow}) \rangle =
\hat{T}_{L-M,\sigma} \ | \Phi_0 (N_{\uparrow \uparrow},N_{\uparrow
\downarrow}) \rangle \eqno(14)
$$
or, in general,
$$
| \Phi_\alpha(N_{\uparrow \uparrow},N_{\uparrow \downarrow}) \rangle =
\hat{T}_{j,\sigma} \ | \Phi_\beta(N_{\uparrow \uparrow},N_{\uparrow
\downarrow}) \rangle \ . \eqno(15)
$$
The dimension ${\Cal D}$ of the subspace ${\Cal
M}(N_{\uparrow},N_{\uparrow \uparrow},N_{\uparrow \downarrow})$ is
$\binom{L}{M}$. Moreover,  these different subspaces are orthogonal and
not mixed by the Hamiltonian $H_{t-J_z}$.

We want to show now that, for a given number of holes $M$, the subspace
generated by the N\'eel parent state, ${\Cal
M}(N_{\uparrow},0,N_{\uparrow \downarrow}) \equiv {\Cal M}_0$, contains
the ground state. To this end, one has to note that the  matrix
elements $\langle \Phi_\alpha(N_{\uparrow \uparrow},N_{\uparrow
\downarrow})| \hat{T} | \Phi_\beta(N_{\uparrow \uparrow},N_{\uparrow
\downarrow}) \rangle$ are the same for the different subspaces ${\Cal
M}$. Nonetheless, the magnetic matrix elements $\langle
\Phi_\alpha(N_{\uparrow \uparrow},N_{\uparrow \downarrow})|
\hat{H}_{J_z} | \Phi_\beta(N_{\uparrow \uparrow},N_{\uparrow
\downarrow}) \rangle = \delta_{\alpha \beta} \ A({\Cal N}_{\uparrow
\uparrow},{\Cal N}_{\uparrow \downarrow})$ are different for the
different subspaces, with $A(0,{\Cal N}_{\uparrow \downarrow}) \leq
A(\bar{{\Cal N}}_{\uparrow \uparrow},\bar{{\Cal N}}_{\uparrow
\downarrow})$ , ${\Cal N}_{\uparrow \downarrow} = \bar{{\Cal
N}}_{\uparrow \uparrow}+\bar{{\Cal N}}_{\uparrow \downarrow}$. [Notice
that, for a generic state of a given subspace, ${\Cal N}_{\uparrow
\uparrow}+ {\Cal N}_{\uparrow \downarrow} \leq  L - M -1$ with the
equality satisfied by the parent state only, where ${\Cal N}_{\uparrow
\uparrow}=N_{\uparrow \uparrow}$ and ${\Cal N}_{\uparrow
\downarrow}=N_{\uparrow \downarrow}$.]  Therefore, the Hamiltonian
matrices H $^{\Cal M}_{\alpha,\beta}$ (of dimension ${\Cal D} \times
{\Cal D}$) in each subspace ${\Cal M}$, consists of identical
off-diagonal matrix elements (H$^{\Cal M}_{\alpha,\beta}$ = H$^{\Cal
M'}_{\alpha,\beta},  \; \alpha \neq \beta$) and different  diagonal
ones. These Hermitian matrices can be ordered according to the
increasing value of the energy $E_M$ of their parent states, which is
equivalent (for fixed $L$ and $M$) to ordering by the increasing number
of ferromagnetic links $N_{\uparrow \uparrow}$, ${\text H}^{\Cal M} 
\equiv {\text H}^{N_{\uparrow \uparrow}}$. For any $N_{\uparrow
\uparrow} < N'_{\uparrow \uparrow}$, ${\text H}^{N'_{\uparrow
\uparrow}} = {\text H}^{N_{\uparrow \uparrow}} + B$, where $B$ is a
positive semidefinite matrix. Then, the monotonicity theorem [8] tells
us that
$$
E_k(N_{\uparrow \uparrow}) \leq E_k(N'_{\uparrow \uparrow}) \;\;\;
\forall \ k=1,\cdots,{\Cal D} \ , \eqno(16)
$$
where $E_k(N_{\uparrow \uparrow})$ are the eigenvalues of ${\text
H}^{N_{\uparrow \uparrow}}$ arranged in increasing order. Therefore, we
conclude that the lowest eigenvalue of $H_{t-J_z}$ must be in ${\Cal
M}_0$  and is $E_1(0)$. 
 
The next step consists in showing that, within the ground state
subspace ${\Cal M}_0$, the Hamiltonian $H_{t-J_z}$ maps into an
attractive spinless fermion model. If one makes the following
identification
$$ 
| \underset{L - M}\to{\underbrace{\uparrow \downarrow \uparrow \downarrow
\cdots}} \ \underset{M}\to{\underbrace{\circ_{\;} \! \!\circ
\circ \circ \cdots}} \ \rangle \rightarrow
| \underset{L - M}\to{\underbrace{\bullet \bullet \bullet \bullet
\cdots}} \ \underset{M}\to{\underbrace{\circ_{\;} \! \!\circ
\circ \circ \cdots}} \ \rangle \ , \eqno(17)
$$
i.e., any spin particle ($c^{\dagger}_{j \sigma}$, independently of the
value of $\sigma$) maps into a single spinless fermion
($b^{\dagger}_{j}$) in ${\Cal M}_0$, it is straightforward to realize
that all matrix elements of H$^0$ are identical to the matrix elements
of
$$ 
H_0 = -t  \sum_{j} (b^{\dagger}_{j} b^{\;}_{j+1} +{\text
H.c.}) - \frac{J_z}{4} \sum_{j} n_{j} n_{j+1} 
\eqno(18)
$$
in the corresponding new basis. In Eq. (18) $n_{j}=b^{\dagger}_{j}
b^{\;}_{j}$.

The addition of the term $\eta \sum_{j \sigma} (\bar{c}^\dagger_{j
\sigma}\bar{c}^\dagger_{j+1 \bar{\sigma}} + {\text {\rm H.c.}})$ to the
$t$-$J_z$ Hamiltonian has two different effects in the lowest energy
subspace ${\Cal M}_0$. The process where the pair of up and down
particles created preserves the N\'eel ordering of the spins can be
mapped into the creation of two spinless particles in the effective
spinless model. The other possible process creates at least one
ferromagnetic link in the parent state and connects ${\Cal M}_0$ with
the subspace containing one ferromagnetic link, $N_{\uparrow
\uparrow}$=1, (the lowest spin excitation) ${\Cal M}_1$. This means
that the subspaces ${\Cal M}$ are no longer invariant under the
application of the Hamiltonian. However this second process contributes
to second order in $\eta$ ($\eta^2/\Delta_s$) due to the existence of a
spin gap between the ground states of ${\Cal M}_0$ and ${\Cal M}_1$ 
(see Appendix I). Therefore to first order in $\eta$, ${\Cal M}_0$ is
still an invariant subspace and the reduced Hamiltonian is a spinless
model with a superconducting term
$$
H'_0 = H_0 +
\eta \sum_{j} (b^{\dagger}_{j}b^{\dagger}_{j+1}+
b^{\;}_{j} b^{\;}_{j+1}) \eqno(19)
$$

For arbitrary values of $J_{z}$, $t$, and hole density $\nu$, $H'_0$ is
equivalent (via the traditional Jordan-Wigner transformation) to the
spin-$\frac{1}{2}$ XYZ chain Hamiltonian (up to an irrelevant constant)
$$ 
H'_0 = \sum_j \left( {\Cal J}_x \ s_{j}^x
s_{j+1}^x + {\Cal J}_y \ s_{j}^y s_{j+1}^y + {\Cal J}_z \
(s_{j}^z s_{j+1}^z+ s_{j}^z) \right) \eqno(20)
$$ 
and ${\Cal J}_x=2 (\eta - t)$, ${\Cal J}_y=-2(\eta+t)$, and ${\Cal
J}_z=-\frac{J_z}{4}$. In the language of our original Hamiltonian
$H_{\text {\rm xxz}}$, Eq. (11), ${\Cal J}_x=2 (\eta + \Delta)$,
${\Cal J}_y=-2(\eta-\Delta)$, and ${\Cal J}_z=-1$. From
exact solution of this model [9], it is seen that the system is
critical only when $\eta=0$ while for $\eta \neq 0$ a gap to all
excitations opens. 
\topinsert
\input psfig.sty
\centerline{\hskip-20mm
\psfig{figure=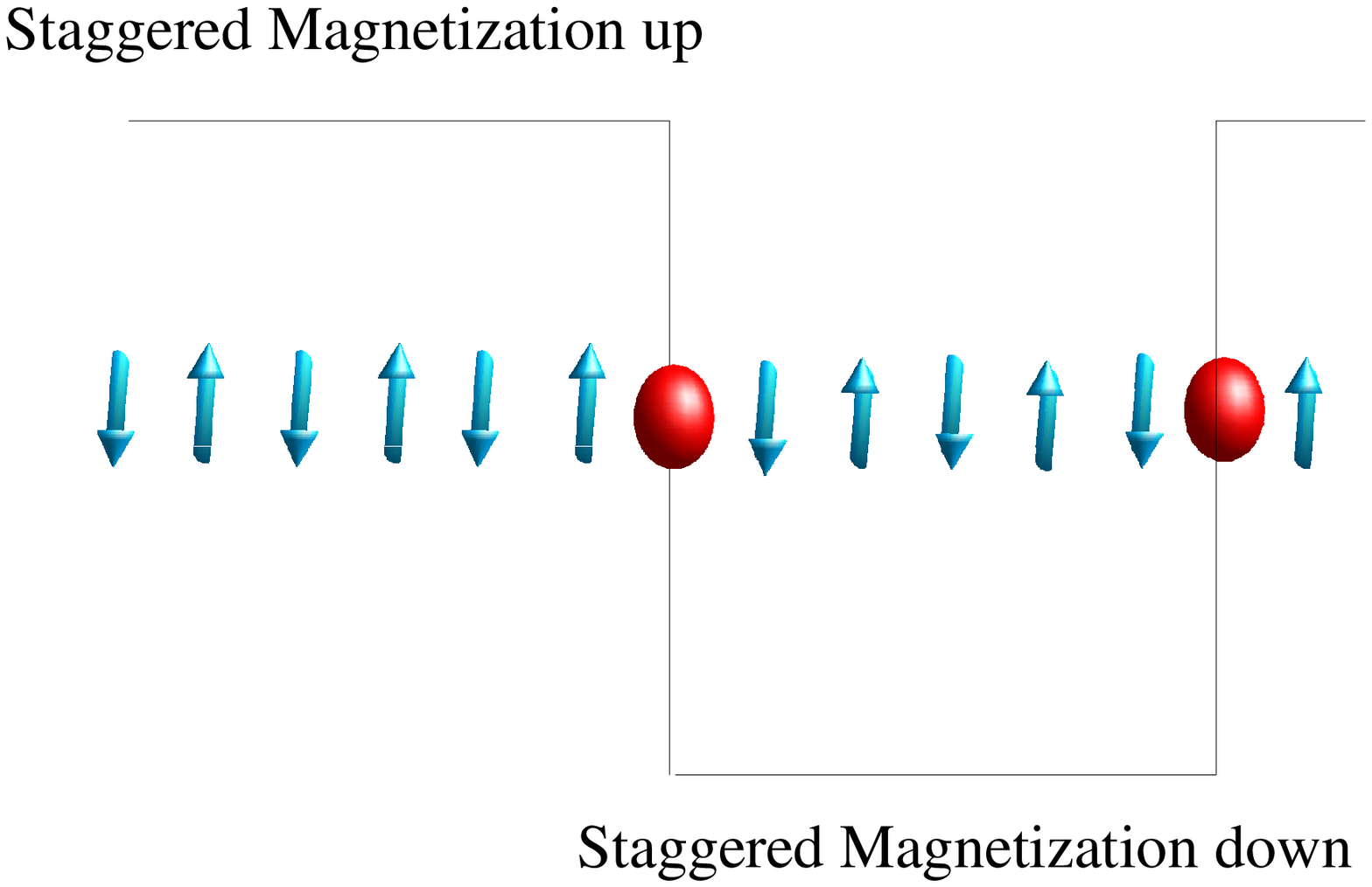,height=7truecm,width=11truecm,angle=0}}
\vskip -0.0truecm 
\noindent
{\bf Figure 2.}
Each hole carries an anti-phase domain wall in the exact ground state of
the $t$-$J_z$ model.  
\vskip 6truept
\endinsert

It is important to note that the ground state of the $t$-$J_z$ model, 
$| \Psi_0^0 \rangle$, has the same topological long range order as the
valence-bond-solid (Haldane state) [10], i.e., the correlation function
[11] $\displaystyle \langle \Psi_0^0 | S^z_j \ \exp[i\pi \hskip-2mm
\sum_{k=j+1}^{j+r-1} \hskip-2mm S^z_k ] \ S^z_{j+r}|\Psi_0^0\rangle= -
\langle \Psi_0^0 |n_j n_{j+r}|\Psi_0^0\rangle$ has a power law decay as
a function of distance $r$ to a constant value for the $t$-$J_z$ point.
This means that the superconducting term in $H_{\text {\rm xxz}}$,
although it is opening a gap (Haldane gap), it is not changing the 
topological order characterizing the ground state. Therefore, the
ground state is in the same Haldane phase for $0 \leq \eta \leq
\Delta$. In the particle language, the Haldane gap is a superconducting
gap. Since each hole ($S^z=0$ state in the spin language) is an
anti-phase boundary (soliton) for the N\'eel ordering (see Fig. (2)),
the AF correlation function is short ranged for the ground state of the
$t$-$J_z$ model. As demonstrated above, these solitonic excitations are
massless at the $t$-$J_z$ point, but become massive (gapped) as soon as
the superconducting term is turned on ($\eta \neq 0$). As the
superconducting term is derived from the  transverse part of the
Heisenberg interaction, it will not restore the AF ordering along the
$z$ direction. In this way it is easy to understand why the spin-spin
correlation function, of the $S$=1 AF Heisenberg chain is short ranged:
$\langle {\text {\rm S}}_j \cdot {\text {\rm S}}_{j+r}\rangle \sim
\exp[-r/\xi]$.

\vskip 28 truept

\centerline{\bf 3.2  $S$=1 Integrable Models and Hidden Symmetries}
\vskip 12 truept

To illustrate further the power of our spin-fermion mapping we now
present {\it exact} solutions of 1$d$ $S$=1 models that have not been
discovered by traditional techniques. These models correspond to the
family of bilinear-biquadratic Hamiltonians 
$$
H_1(\Delta) = \sum_j H_j^z + H_j^{\text {\rm xx}} + \left \{ H_j^z,
H_j^{\text {\rm xx}} \right \} = \sum_j H_j^z + \Delta \sum_\sigma
\bar{c}^\dagger_{j \sigma} \bar{c}^{\;}_{j+1 \sigma} \ , \eqno(21)
$$
that can be mapped onto a ($S=$$\frac{1}{2}$) $t$-$J_z$ Hamiltonian,
whose quantum phase diagram has recently been exactly solved [7].

Another well-studied class of bilinear-biquadratic $SU(2)$ invariant
Hamiltonians is [12]
$$
H_2(\Delta) = \sum_j {\text {\bf S}}_j \cdot {\text {\bf S}}_{j+1} +
\Delta \left ( {\text {\bf S}}_j \cdot {\text {\bf S}}_{j+1} \right )^2
\ , \eqno(22)
$$
where -1$\leq$ $\Delta$ $\leq$1. The pure Heisenberg ($\Delta$=0) and
Valence Bond Solid models ($\Delta=\frac{1}{3}$) belong to the Haldane
gapped phase, that extends over the whole interval of $\Delta$ except
at the boundaries $\Delta=\pm 1$ which are quantum critical points. The
case $\Delta$= -1 is known to be Bethe ansatz solvable with a unique
ground state and gapless. For $\Delta$=1 we can map $H_2(1)$ onto the
supersymmetric ($S$=$\frac{1}{2}$) $t$-$J$ Hamiltonian plus a
NN repulsive interaction 
$$
H_2(1)=-\sum_{j \sigma} \left ( \bar{c}^\dagger_{j \sigma}
\bar{c}^{\;}_{j+1 \sigma} + {\text {\rm H.c.}} \right ) + 2 \sum_j
{\text {\bf s}}_j \cdot {\text {\bf s}}_{j+1} + 2 \sum_j (1-\bar{n}_j +
\frac{3}{4} \bar{n}_j \bar{n}_{j+1}) \ , \eqno(23)
$$
where ${\text {\bf s}}_j$ represents a $S$=$\frac{1}{2}$ operator. This
model is Bethe-ansatz solvable with a gapless phase [13] and is known as
the Lai-Sutherland solution [14].

We now discuss the importance of our generalized Jordan-Wigner
transformation in unraveling hidden symmetries of an arbitrary spin
Hamiltonian. In Eq. (22), for example, the $S$=1 $SU(2)$ symmetry
is manifest. However, both the $S$=$\frac{1}{2}$ $SU(2)$ and global
$U(1)$ gauge symmetries are hidden. On the other hand, in the
transformed Hamiltonian, Eq. (23), these two symmetries are
manifested explicitly through rotational invariance and charge
conservation. The generators of these symmetries are related through
the mapping already introduced. To illustrate this, we consider the
$U(1)$ symmetry case. Here the generator of the transformation is
$Q=\sum_j \bar{n}_{j}$ which maps onto $Q=\sum_j (S_j^z)^2$ in the spin
representation. The total group symmetry of the Hamiltonian is $SU(3)$. 
To see this explicitly, let us rewrite Eq. (23) in a way that is
explicitly $SU(3)$ invariant (up to an irrelevant constant) [16]
$$
H_2(1)= \sum_j {\Cal S}^{\mu \nu}(j) {\Cal S}_{\nu \mu}(j+1) \ ,
$$
where ${\Cal S}^{\mu \nu}(j)$ is a nine component tensor (traceless,
i.e., Tr$[{\Cal S}]=0$)
$$
{\Cal S}(j)= \pmatrix{\bar{n}_{j \uparrow}
-\frac{1}{3}&\bar{c}^\dagger_{j \uparrow} K_j& \bar{c}^\dagger_{j
\downarrow} K_j \cr  &&\cr K^\dagger_j \bar{c}^{\;}_{j \uparrow}&\bar{n}_{j
\downarrow}-\frac{1}{3}& \bar{c}^\dagger_{j \uparrow} \bar{c}^{\;}_{j
\downarrow} \cr &&\cr K^\dagger_j \bar{c}^{\;}_{j
\downarrow}&\bar{c}^\dagger_{j \downarrow} \bar{c}^{\;}_{j
\uparrow}&-(\bar{n}_{j \uparrow}+\bar{n}_{j \downarrow}-\frac{2}{3})
\cr}
\ ,
$$
whose eight components (${\Cal S}^{33}(j)$ is linearly dependent)
constitute a basis of the $su(3)$ algebra [16].

\vskip 28 truept

\centerline{\bf 4. GENERALIZED TRANSFORMATIONS}
\vskip 12 truept

A general transformation for arbitrary spin and spatial dimension is
the following (see Fig. (3))

\noindent
{\bf Half-odd integer spin $S$} ($\sigma \in {\Cal F}_{1 \over 2} = 
\{-S+1, \dots, S \}$): 
$$
\eqalignno{
S^+_j &= \eta_{\bar{S}} \ \bar{c}^\dagger_{j \bar{S}+1} \ K_j + \sum
\Sb \sigma \in {\Cal F}_{\frac{1}{2}} \\ \sigma \neq S \endSb \!
\eta_\sigma \ \bar{c}^\dagger_{j \sigma+1} \bar{c}^{\;}_{j \sigma}
\ , \cr
S^-_j &= \eta_{\bar{S}} \ K_j^\dagger \ \bar{c}^{\;}_{j \bar{S}+1} +
\sum \Sb \sigma \in {\Cal F}_{\frac{1}{2}} \\ \sigma \neq S \endSb \!
\eta_\sigma \ \bar{c}^\dagger_{j \sigma} \ \bar{c}^{\;}_{j \sigma+1}
\ , \cr
S^z_j &= -S + \sum \Sb \sigma \in {\Cal F}_{\frac{1}{2}} \endSb
(S+\sigma) \ \bar{n}_{j \sigma} \ , \cr
\bar{c}^\dagger_{j \sigma} &= K_j^\dagger L_\sigma^{\frac{1}{2}} \left( S^+_j
\right)^{\sigma+S} {\Cal P}_j^{\frac{1}{2}} \ , \
{\text {\rm where}} \  {\Cal P}_j^{\frac{1}{2}} =
\!\! \prod \Sb \tau \in {\Cal F}_{\frac{1}{2}} \endSb
\! \frac{\tau -S^z_j}{\tau + S} \ , \ L_\sigma^{\frac{1}{2}} =
\!\!\!\!\! \prod_{\tau=-S}^{\sigma-1} \!\!\! \eta_\tau^{-1} \ . 
&(24)}
$$

\noindent
{\bf Integer spin $S$} ($\sigma \in {\Cal F}_1 =
\{-S,\dots,-1,1,\dots,S \}$): 
$$
\eqalignno{
S^+_j &= \eta_0 \ (\bar{c}^\dagger_{j 1} \ K_j + K_j^\dagger \
\bar{c}^{\;}_{j \bar{1}}) +  \sum \Sb \sigma \in {\Cal F}_1 \\ \sigma \neq
-1,S \endSb \! \eta_\sigma \ \bar{c}^\dagger_{j \sigma+1}
\bar{c}^{\;}_{j \sigma} \ ,  \cr 
S^-_j &= \eta_0 \ (K_j^\dagger \ \bar{c}^{\;}_{j 1} +
\bar{c}^\dagger_{j \bar{1}} \ K_j) + \sum \Sb \sigma \in {\Cal F}_1 \\ \sigma
\neq -1,S \endSb \! \eta_\sigma \ \bar{c}^\dagger_{j \sigma} \
\bar{c}^{\;}_{j \sigma+1} \ , \cr 
S^z_j &= \sum \Sb \sigma \in {\Cal F}_1 \endSb \sigma \
\bar{n}_{j \sigma} \ , \cr
\bar{c}^\dagger_{j \sigma} &= K_j^\dagger L_\sigma^1 \ 
            \cases{
            \left( S^+_j \right)^\sigma {\Cal P}_j^1 \ ,& 
            {\text {\rm if $\sigma > 0$}}   \cr
            \left( S^-_j \right)^\sigma {\Cal P}_j^1 \ ,& 
            {\text {\rm if $\sigma < 0$}} \ , 
            \cr }
	    &(25)}
$$
where $\displaystyle {\Cal P}_j^1 = \!\! \prod \Sb \tau \in {\Cal F}_1
\endSb \! \frac{\tau -S^z_j}{\tau} \ , \ L_\sigma^1 = \!
\prod_{\tau=0}^{|\sigma|-1}  \eta_\tau^{-1}$ and $\eta_\sigma =
\sqrt{(S-\sigma)(S+\sigma+1)}$.
	    
The total number of flavors is $N_f=2 S$, and the $S$=$\frac{1}{2}$
case simply reduces to the traditional Jordan-Wigner transformation.
Since these mappings are exact they preserve the invariant Casimir
operator ${\text {\bf S}}_j^2 = S(S+1)$. The generalized constrained fields 
$$
\bar{c}^\dagger_{j \sigma} = {c}^\dagger_{j \sigma} \!\!\! \prod \Sb
\tau \in {\Cal F}_\alpha \endSb \!\! (1-{n}_{j
\tau}) \ , \ 
\bar{c}^{\;}_{j \sigma} = \!\!\!\! \prod \Sb \tau \in {\Cal F}_\alpha
\endSb  \!\!(1-{n}_{j \tau}) \ {c}^{\;}_{j \sigma}
\ \eqno(26)
$$
form a subalgebra of the generalized Hubbard double graded algebra,
where the ``unconstrained'' operators $c^\dagger_{j \sigma}, c^{\;}_{j
\sigma}$ satisfy the standard fermion anticommutation relations 
($\alpha=\frac{1}{2},1$ depending upon the spin character of the
representation). These generalized constrained operators (only single
occupancy is allowed) anticommute for different sites
$$
\eqalignno{
\left \{ \bar{c}^{\;}_{j \sigma}, \bar{c}^{\;}_{k \sigma'} \right \} &=
\left \{ \bar{c}^\dagger_{j \sigma}, \bar{c}^\dagger_{k \sigma'} \right
\} = 0  \ , \
\left \{ \bar{c}^{\;}_{j \sigma}, \bar{c}^\dagger_{k \sigma'} \right \}
= \delta_{jk} \cases{
\hbox{$\displaystyle \prod \Sb \tau \in {\Cal F}_\alpha  \cr \tau
                \neq \sigma \endSb  \!\!(1-\bar{n}_{j \tau})$} \ ,&
                {\text {if $\sigma = \sigma'$}} , \cr
                \bar{c}^\dagger_{j \sigma'} \bar{c}^{\;}_{j \sigma} \ ,& 
                {\text {if $\sigma \neq \sigma'$}} , 
                \cr} 
&(27)}
$$
and their number operators satisfy $\bar{n}_{j \sigma}\bar{n}_{j
\sigma'}= \delta_{\sigma \sigma'} \bar{n}_{j \sigma}$.

The string operators $K_j$ introduce nonlinear and nonlocal
interactions between the constrained fermions. For 1$d$ lattices ($K_j
= K_j^\dagger$, $\left[K_i, K_j\right]=0$) they are the so-called kink
operators $K_j= \exp[i \pi \sum\limits_{k < j} \bar{n}_k]$,
while for 2$d$ [15]
$$
K_{\text {\bf j}} = \exp[i \sum \Sb {\text {\bf k}} \endSb a({\text {\bf
k}}, {\text {\bf j}})  \ \bar{n}_{\text {\bf k}}] \ , \;\; \hbox{with} \ 
\bar{n}_{\text {\bf k}} = \sum \Sb \sigma \in {\Cal F}_\alpha \endSb 
\bar{n}_{{\text {\bf k}} \sigma} = 1- {\Cal P}_{\text {\bf k}}^\alpha \
. \eqno(28)
$$
Here, $a({\text {\bf k}},{\text {\bf j}})$ is the angle between the
spatial vector ${\text {\bf k}}-{\text {\bf j}}$ and a fixed direction
on the lattice, and $a({\text {\bf j}},{\text {\bf j}})$ is defined to
be zero. We comment that the 1$d$ kink operators constitute a
particular case of Eq. (28) with $a(k,j)=\pi$ when $k<j$ and equals
zero otherwise. For $d>2$, the string operators generalize [16] along
the lines introduced in Ref. [17]. 
\topinsert
\input psfig.sty
\centerline{\hskip0mm
\psfig{figure=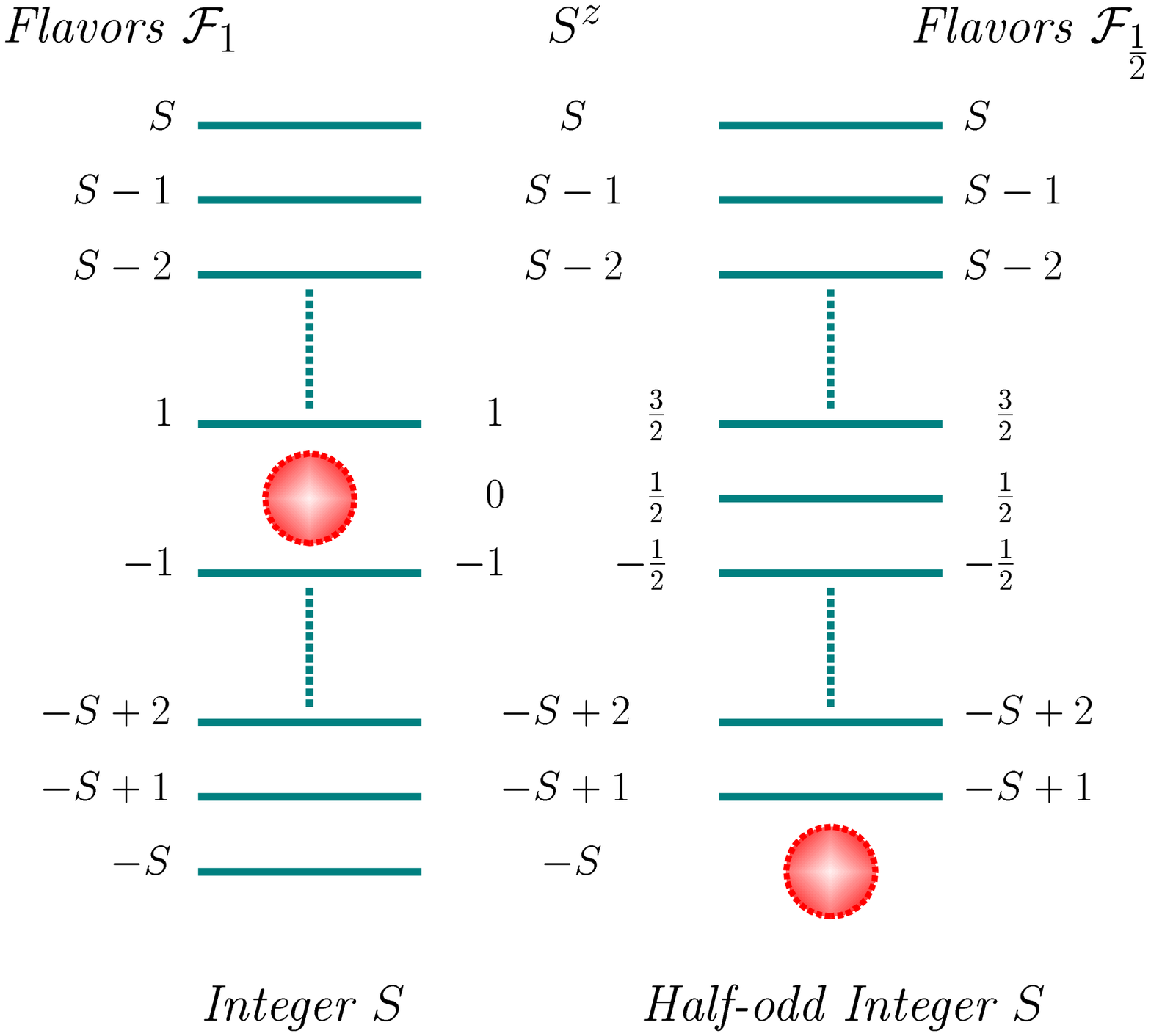,height=9truecm,width=10truecm,angle=0}}
\vskip -0.0truecm 
\noindent
{\bf Figure 3.} 
Constrained fermion states per site for integer and half-odd
integer spin $S$. In both cases there are $2S$ flavors and the 
corresponding $2S+1$ values of $S^z$ are shown in the middle column.
One degree of freedom is assigned to the fermion vacuum (circle) whose
relative position depends upon the spin character.\vskip 12truept
\vskip 6truept
\endinsert

Why are the vacuum states chosen like in Fig. (3)? Note that from a
mathematical viewpoint the vacuum could be identified with any
$S^z$ value. However, our choice is the most symmetric one and
therefore the most useful one to establish connections between
Hamiltonians which are relevant in different fields of physics. The
identification of the vacuum state with $S^z=0$ can only be done in the
integer case.

There is always the freedom to perform rotations in spin space to get
equivalent representations to the one presented above. However, for
bilinear isotropic NN Heisenberg (spin $SU(2)$ rotationally invariant)
Hamiltonians in the large-$S$ limit, there is a fundamental difference
between effective integer and half-odd integer spin cases. In the
latter case a new local $U(1)$ gauge symmetry emerges that is
explicitly broken in the integer case. For 1$d$ lattices, this is
precisely what distinguishes Haldane gap systems [6] from half-odd
integer spin chains that are critical. 

We mention that other fermionic representations are feasible. In
particular, for half-odd integer cases where $2S+1 =
\sum_{i=0}^{\bar{N}_f} {\binom{\bar{N}_f}{i}}= 2^{\bar{N}_f}$ (e.g.,
$S$=$\frac{3}{2}$ with $\bar{N}_f=2$) a simple transformation in terms
of standard ``unconstrained'' fermions is possible [16].  For these
mappings the string operators must be modified to take into account the
double occupancy of a site. In this way the Hubbard model can be mapped
onto a $S=\frac{3}{2}$ spin Hamiltonian [16].

\vskip 28 truept

\centerline{\bf 5. TWO DIMENSIONAL LATTICES AND SPIN-ANYON MAPPING}
\vskip 12 truept

The generalization of these transformations to higher dimensions gives
new exact mappings between spin theories and constrained fermion
systems in the presence of gauge fields. To illustrate this we write
the $S=$1 Hamiltonian $H_2(1)$ in the fermion representation for $d=2$ 
$$
H_2(1)=- \! \sum_{{\text {\bf j}} \sigma,\nu} \left ( \bar{c}^\dagger_{
{\text {\bf j}}+{\text {\bf e}}_\nu \sigma} \  e^{iA_{\nu}({\text {\bf
j}} )} \ \bar{c}^{\;}_{{\text {\bf j}} \sigma} + {\text {\rm H.c.}} \right )
+ 2 \sum_{{\text {\bf j}},\nu} {\text {\bf s}}_{{\text {\bf j}}} \cdot
{\text {\bf s}}_{{\text {\bf j}}+{\text {\bf e}}_\nu} 
$$
$$ 
\hskip16mm + \sum_{{\text {\bf j}}, \nu} \left (2-(\bar{n}_{{\text {\bf
j}}}+\bar{n}_{{\text {\bf j}}+{\text {\bf e}}_\nu}) + \frac{3}{2}
\bar{n}_{{\text {\bf j}}} \bar{n}_{{\text {\bf j}}+{\text {\bf e}}_\nu}
\right ) \ , \eqno(29)
$$
and
$$
A_{\nu}({\text {\bf j}})=\sum_{{\text {\bf k}}}[a({\text {\bf
k}},{\text {\bf j}})-a({\text {\bf k}},{\text {\bf j}}+{\text {\bf
e}}_\nu)] \ \bar{n}_{\text {\bf k}} \ , \eqno(30)
$$
where ${\text {\bf e}}_\nu$ ($\nu=1,2$) are basis vectors of the
Bravais lattice connecting NN and ${\text {\bf j}}$'s represent sites
of the corresponding 2$d$ lattice. We note that the field
$A_{\nu}({\text {\bf j}})$ is associated with the change in particle
statistics. It is well-known [15,3] that the same transmutation of
particle statistics can be achieved via a  path-integral formulation
for $H_2(1)$ where an Abelian lattice Chern-Simons term is included. In
this formulation a constraint (Gauss's law) requiring that the gauge
flux through a plaquette ${\text {\bf j}}$ be proportional to the total
fermion density on the site, $\bar{n}_{\text {\bf j}}$, is enforced. 
This suggests that our spin-fermion mapping can be generalized to an
spin-anyon transformation with a hard-core condition for the anyon
fields [16]. In fact, one can formally take our generalized
Jordan-Wigner transformation and replace the string operators $K_{\text
{\bf j}}$ by the statistical operators $K_{\text {\bf j}}(\theta) =
\exp[i \theta \sum \Sb {\text {\bf k}} \endSb a({\text {\bf k}},{\text
{\bf j}}) \bar{n}_{\text {\bf k}}]$ with $0 \leq \theta \leq 1$. With
this choice, the $\bar{c}$ operators satisfy equal-time anyon
commutation relations [$\theta =1(0)$ corresponds to constrained
fermions(bosons)] [16]. Similar ideas apply for 1$d$ lattices. 

One immediately sees the relevance of these transformations for the
theories of magnetism and high-temperature superconductivity: A class
of $S=$1 Hamiltonians that can be mapped onto a lattice-gauge
(Chern-Simons) $S=\frac{1}{2}$ $t$-$J$ theory and vice versa; for
example, a $S=\frac{1}{2}$ $t$-$J$ model,  
$$
H_{t-J} =-t\sum_{{\text {\bf j}} \sigma,\nu} \left (
\bar{c}^\dagger_{{\text {\bf j}} \sigma} \bar{c}^{\;}_{{\text {\bf
j}}+{\text {\bf e}}_\nu \sigma} + {\text {\rm H.c.}} \right ) + J
\sum_{{\text {\bf j}},\nu} {\text {\bf s}}_{\text {\bf j}} \cdot {\text
{\bf s}}_{{\text {\bf j}}+{\text {\bf e}}_\nu} - \mu \sum_{\text {\bf
j}} \bar{n}_{\text {\bf j}}  \ , \eqno(31)
$$
can be exactly mapped onto a lattice-gauge bilinear-biquadratic $S=$1
theory
$$
H_{t-J} = 
\frac{J}{8}\sum_{{\text {\bf j}},\nu} \left ( H^z_{{\text {\bf j}} \nu}
\! - \! \frac{4t}{J} S^+_{\text {\bf j}} e^{iA_{\nu}({\text {\bf j}})}
S^-_{{\text {\bf j}}+{\text {\bf e}}_\nu} - \frac{4t}{J} \left \{
H^z_{{\text {\bf j}} \nu}, S^+_{\text {\bf j}} e^{iA_{\nu}({\text {\bf
j}})} S^-_{{\text {\bf j}}+{\text {\bf e}}_\nu}\right \} \right.
$$
$$
\hskip49mm + \left .
(S^+_{\text {\bf j}} S^-_{{\text {\bf j}}+{\text {\bf e}}_\nu} )^2  +
{\text {\rm H.c.}} \right ) -\mu \sum_{{\text {\bf j}}} (S^z_{\text {\bf
j}})^2 \ . \eqno(32)
$$

By means of a semiclassical approximation it has been shown [18] that
the ground state of $H_2(1)$ is on the  boundary between AF ($\Delta <
1$) and orthogonal nematic (non-uniform, $\Delta > 1$) phases [18,12].
These two states are the result of the competition between the
quadratic and quartic spin-exchange interactions in $H_2(\Delta)$. In
terms of the equivalent $t$-$J$ gauge theory this translates into a
competition between antiferromagnetism and delocalization.
Qualitatively, the string-path of the particle moving in an AF
background gives rise to a linear confining potential since the number
of frustrated magnetic links is proportional to the length of the path.
This observation suggests that the inhomogeneous phases observed in the
``striped'' high-T$_c$ compounds can be driven by the competition
between magnetism and delocalization.

\vskip 28 truept

\centerline{\bf 6. SUMMARY}
\vskip 12 truept

We introduced a general spin-particle mapping for arbitrary spin $S$
and spatial dimension that naturally generalizes the Jordan-Wigner
transformation for $S=\frac{1}{2}$. Indeed our transformations define
exact connections between localized quantum spins $S$ on a lattice and
quantum lattice gases of itinerant particles with ``effective''
spin $s=S-\frac{1}{2}$. Mathematically, we established a one-to-one
mapping of elements of a Lie algebra onto elements of a fermionic
algebra with a hard-core constraint. Several generalizations, like a
spin-anyon mapping, and important consequences result from these
transformations [16]. Incidentally, we note that there are extremely
powerful numerical techniques (cluster algorithms [19]) to study
quantum spin systems, and our mapping allows one to extend these
methods to study the equivalent fermionic problems. Generalizations of
these  spin-particle transformations to $SU(N)$ algebras exist [16]. As
it has been illustrated in a previous Section this type of mappings is
very useful to find particle models having internal symmetries which
simplify the resolution of those  problems. In this way it is possible
to find exact solutions in any dimension having more than one
spontaneously broken  symmetry. For instance, we have found exact
solutions in any spatial dimension for $S=\frac{1}{2}$ hard-core bosons
which are magnetic Bose-Einstein condensates.  On the other hand, the
existence of a Quantum Link Model connecting Lattice-Gauge theories to
spin theories in addition to the transformations introduced in this
paper, opens the possibility of formal connections between gauge
theories of high-energy physics and strongly correlated problems of
condensed matter. Finally, it is important to mention that there are
other possible spin-particle mappings such as connections between spins
and canonical fermions. However, the latter transformations are only
feasible for some irreducible spin representations.

\vskip 28 truept

\centerline{\bf ACKNOWLEDGMENTS}
\vskip 12 truept

We thank J.E. Gubernatis for a careful reading of the manuscript and 
G.A. Baker, Jr. for helping us with TeX. This work was sponsored by the
US DOE under contract W-7405-ENG-36.

\vskip 28 truept

\centerline{\bf APPENDIX I}
\vskip 12 truept

\noindent
In this Appendix we prove that the 1$d$ $t$-$J_z$ model has a gap to
spin excitations. We follow the notation of Section (3.1). 

The spin gap is defined as
$$ 
\Delta_s= \langle\Psi_0^1|H_0^1|\Psi_0^1\rangle - 
\langle\Psi_0^0|H_0^0|\Psi_0^0\rangle \geq 0 \ , \eqno(A1)
$$
where $H_0^0$ is the Hamiltonian restricted to the ${\Cal M}_0$
subspace, while $H_0^1$ is the one restricted to ${\Cal M}_1$ (${\Cal
M}_1$ is the subspace generated from a N\'eel ordered parent state with
$N_{\uparrow \uparrow}=1$)
$$ 
{\Cal M}_0: \ | \underset{L - M}\to{\underbrace{\uparrow \downarrow
\uparrow \downarrow \cdots {\buildrel \nu \over \uparrow} \downarrow
\uparrow \downarrow \uparrow \cdots}} \ \underset{M}\to{\underbrace{\circ_{\;}
\! \!\circ \circ \circ \cdots}} \ \rangle \ \hbox{and } \
{\Cal M}_1: \ | \underset{L - M}\to{\underbrace{\uparrow \downarrow
\uparrow \downarrow \cdots {\buildrel \nu \over \uparrow} \uparrow \downarrow
\uparrow \downarrow \cdots}} \ \underset{M}\to{\underbrace{\circ_{\;}
\! \!\circ \circ \circ \cdots}} \ \rangle 
\ . 
$$
$|\Psi_0^0\rangle$ and $|\Psi_0^1\rangle$ are the normalized ground
states ($\langle \Psi_0^i|\Psi_0^i\rangle$=1) of $H_0^0$ and $H_0^1$,
respectively. From the Rayleigh-Ritz variational principle
$$ 
\Delta_s \geq \langle\Psi_0^1|H_0^1|\Psi_0^1\rangle - 
\langle\Psi_0^1|H_0^0|\Psi_0^1\rangle =
\langle\Psi_0^1|H_0^1-H_0^0|\Psi_0^1\rangle \ . \eqno(A2)
$$
Since $H_0^1-H_0^0 = \frac{J_z}{2} \ n_{\nu} n_{\nu+1}$ we have to show
that the correlation function $\tilde{\Delta}= \langle\Psi_0^1|n_{\nu}
n_{\nu+1}|\Psi_0^1\rangle$ is finite in order to demonstrate that there
is a finite spin gap $\Delta_s$. 

The wave function $|\Psi_0^1\rangle$ can be written as
$$
|\Psi_0^1\rangle=\hskip-5mm \sum_{i_1<i_2< \cdots <i_{L-M}} \hskip-5mm
a_{I} \ b^{\dagger}_{i_1} b^{\dagger}_{i_2} \cdots
b^{\dagger}_{i_{\nu}} b^{\dagger}_{i_{\nu+1}} \cdots
b^{\dagger}_{i_{L-M}} |0 \rangle \eqno(A3)
$$
in the basis $\{ b^{\dagger}_{i_1} b^{\dagger}_{i_2} \cdots
b^{\dagger}_{i_{\nu}} b^{\dagger}_{i_{\nu+1}} \cdots
b^{\dagger}_{i_{L-M}} |0 \rangle \}$ with ${I}=
(i_1,i_2,\cdots,i_{L-M})$ and $i_\mu \in [1,L]$. [$I$ represents a
generic particle configuration. Note that $i_\nu=\nu$ only for the
parent state particle configuration.] If one assumes that
$\tilde{\Delta}=0 ({\Cal O}(1/L))$, then $|\Psi_0^1\rangle$ has zero
(or infinitesimal) projection on the subspace of states where
$i_{\nu}+1=i_{\nu+1}$. This means that there must be at least one empty
site between particles $\nu$ and $\nu+1$ in all the terms of
$|\Psi_0^1\rangle$, i.e., $i_{\nu}+\alpha=i_{\nu+1}$ with $\alpha>1$
(the rest is a set of zero measure). Different values of $\alpha$
define different subspaces of the Hilbert space: $\{ b^{\dagger}_{i_1}
b^{\dagger}_{i_2} \cdots b^{\dagger}_{i_{\nu}}
b^{\dagger}_{i_{\nu}+\alpha} \cdots b^{\dagger}_{i_{L-M}} |0\rangle\}$.
For finite concentration of  particles there is a minimum value
$\alpha_m$ for which the wave function $|\Psi_0^1\rangle$ has a finite
projection onto the corresponding subspace. We call the projected wave
function $|\Psi_p\rangle= P_{\alpha_m}|\Psi_0^1\rangle$ with 
$$
P_{\alpha}=\hskip-5mm \sum \Sb i_1<i_2<\cdots<i_{L-M}, \\
i_{\nu}+\alpha=i_{\nu+1} \endSb \hskip-5mm n_{i_1} n_{i_2} \cdots 
n_{i_{L-M}} \eqno(A4)
$$
with $n_{i_\mu}=b^{\dagger}_{i_\mu}b^{\;}_{i_\mu}$. We define now the
state
$$
\sqrt{\langle\Psi_p|\Psi_p\rangle} \ |\Psi_1\rangle={\hat{\Cal 
O}}_{\nu,\alpha_m}|\Psi^1_0\rangle = \hskip-5mm \sum \Sb i_1<i_2< \cdots
<i_{L-M} , \cr i_{\nu}+\alpha_m=i_{\nu+1} \endSb \hskip-5mm a_{I} \
b^{\dagger}_{i_1} b^{\dagger}_{i_2}\cdots
b^{\dagger}_{i_{\nu}+1}b^{\dagger}_{i_{\nu+1}} b^{\dagger}_{i_{L-M}}
|0\rangle  \eqno(A5)
$$
with
$$
{\hat{\Cal O}}_{\nu,\alpha}=\hskip-5mm \sum \Sb i_1<i_2<\cdots<i_{L-M}, \cr
i_{\nu}+\alpha=i_{\nu+1} \endSb \hskip-5mm n_{i_1} n_{i_2} \cdots
b^{\dagger}_{i_{\nu}+1} b^{\;}_{i_{\nu}} n_{i_{\nu+1}} \cdots
n_{i_{L-M}} \eqno(A6)
$$
$|\Psi_1\rangle$ is a normalized state which is orthogonal to
$|\Psi_0^1\rangle$  because it belongs to the subspace $\alpha_m-1$. 


One can show that $H_0^1$ connects the states
$|\Psi_0^1\rangle$ and $|\Psi_1\rangle$ with the result 
$$
\langle\Psi_1|H_0^1|\Psi_0^1\rangle = -2t \langle\Psi_p|\Psi_p\rangle +
 {\Cal O}(1/L) > 0 \eqno(A7)
$$ 
One can also compute $\langle\Psi_1|H_0^1|\Psi_1\rangle$
$$
\langle\Psi_1|H_0^1|\Psi_1\rangle = \frac {\langle\Psi^1_0|{\hat {\Cal
O}}^{\dagger}_{\nu,\alpha_m} H_0^1  {\hat {\Cal
O}}_{\nu,\alpha_m}|\Psi^1_0\rangle}{\langle\Psi_p|\Psi_p\rangle} \ .
\eqno(A8)
$$ 
Then,
$$
\langle\Psi_1|H_0^1|\Psi_1\rangle = \frac{\langle\Psi_0^1|{\hat{\Cal
O}}^{\dagger}_{\nu,\alpha_m}  {\hat{\Cal O}}_{\nu,\alpha_m}
H_0^1|\Psi_0^1\rangle}{\langle\Psi_p|  \Psi_p\rangle}  + \frac{
\langle\Psi_0^1|\hat{\Cal O}^{\dagger}_{\nu,\alpha_m}[H_{0}^1,{\hat
{\Cal O}}_{\nu,\alpha_m}]|\Psi_0^1\rangle}{\langle\Psi_p| 
\Psi_p\rangle} \ , \eqno(A9)
$$
and, therefore
$$
\langle\Psi_1|H_0^1|\Psi_1\rangle =  E_0^1+
\frac{\langle\Psi_0^1|{\hat{\Cal
O}}^{\dagger}_{\nu,\alpha_m}[H_{0}^1,{\hat {\Cal
O}}_{\nu,\alpha_m}]|\Psi_0^1\rangle}{\langle\Psi_p|\Psi_p\rangle}  \ ,
\eqno(A10)
$$
where we have used that $|\Psi_0^1\rangle$ is the ground state of
$H_0^1$, $H_0^1|\Psi_0^1\rangle=E_0^1|\Psi_0^1\rangle$ and
$\langle\Psi_0^1|{\hat {\Cal O}}^{\dagger}_{\nu,\alpha_m} {\hat {\Cal
O}}_{\nu,\alpha_m}|\Psi_0^1\rangle=\langle\Psi_p|\Psi_p\rangle$.  The
second term in Eq. (A10) is bounded (${\Cal O}(1)$) since 
$[H_{0}^1,{\hat {\Cal O}}_{\nu,\alpha}]$ is an operator which only
affects the particles $i_{\nu}$ and $i_{\nu+1}$. We can then build a
state which is a linear combination of $|\Psi_0^1\rangle$ and
$|\Psi_1\rangle$ having an energy lower than $E_0^1$ because $H_0^1$
connects the two states. But this contradicts the initial hypothesis
which stated that $|\Psi_0^1\rangle$ was the ground state of $H_0^1$.
Therefore, $\tilde{\Delta}$ is finite and there is a spin gap. 

\vskip 28 truept

\centerline{\bf REFERENCES}
\vskip 12 truept


\item{[1]}
P. Jordan and E. Wigner, Z. Phys. {\bf 47}, 631 (1928).

\item{[2]} 
For a $d$-dimensional lattice with $L$ sites (modes), the operator
$S_j^\mu$ is defined in terms of a Kronecker product $\otimes$ as:
$S_j^\mu = \one \otimes \one \otimes \cdots \otimes 
\underbrace{S^\mu}_{j^{th}} \otimes \cdots \otimes \one \ ,$ where
$\one$ is the $(2S+1) \times (2S+1)$ unit matrix and $S^\mu$ is a 
spin-$S$ operator. Thus $S_j^\mu$ admits a matrix representation of
dimension $(2S+1)^L \times (2S+1)^L$.

\item{[3]}
A.M. Tsvelik, {\it Quantum Field Theory in Condensed Matter Physics}
(Cambridge, Cambridge, 1995).

\item{[4]}
In this particular case, instead of $f^\dagger_j = \frac{1}{\sqrt{2}} \
\exp[i \pi \sum_{k<j} (S_k^z)^2] \  S_j^+$, we could have used
$f^\dagger_j = \frac{1}{\sqrt{2}} \ \exp[i \pi \sum_{k<j} S_k^z] \ 
S_j^+$. We thank Dan Arovas for this remark.

\item{[5]}
E.H. Lieb, T.D. Schultz, and D.C. Mattis, Ann. Phys. {\bf 16}, 407 (1961);
I. Affleck and E.H. Lieb, Lett. Math. Phys. {\bf 12}, 57 (1986).
 
\item{[6]}
F.D.M. Haldane, Phys. Lett. A {\bf 93}, 464 (1983).

\item{[7]} 
C.D. Batista and G. Ortiz, Phys. Rev. Lett. (in press);
cond-mat/0003158. 

\item{[8]}  
R.A. Horn and C.R. Johnson, {\it Matrix Analysis} (Cambridge, New York,
1999).

\item{[9]}
R.J. Baxter, {\it Exactly Solved Models in Statistical Mechanics}
(Academic Press, London, 1990). 

\item{[10]}
I. Affleck, T. Kennedy, E. H. Lieb, and H. Tasaki, Phys. Rev. Lett. {\bf
59}, 799 (1987).

\item{[11]}
M. Den Nijs and K. Rommelse, Phys. Rev. B {\bf 40}, 4709 (1989);
S.M. Girvin and D.P. Arovas, Phys. Scr. T {\bf 27}, 156 (1989);
Y. Hatsugai and M. Kohmoto, Phys. Rev. B {\bf 44}, 11789 (1991);
K. Hida, Phys. Rev. B {\bf 45}, 2207 (1991);
Y. Hatsugai, J. Phys. Soc. Jpn. {\bf 61}, 3856 (1992);
M. Kohmoto and H. Tasaki, Phys. Rev. B {\bf 46}, 3486 (1992);
F.C. Alacaraz and Y. Hatsugai, Phys. Rev. B {\bf 46}, 13914 (1992).

\item{[12]}
G. F\'ath and J. S\'olyom, Phys. Rev. B {\bf 51}, 3620 (1995).

\item{[13]}
P. Schlottmann, Phys. Rev. B {\bf 36}, 5177 (1987). 

\item{[14]}
C.K. Lai, J. Math. Phys. {\bf 15}, 1675 (1974); B. Sutherland, Phys.
Rev. B {\bf 12}, 3795 (1975). 

\item{[15]}
E. Fradkin, Phys. Rev. Lett. {\bf 63}, 322 (1989); D. Eliezer and G.W.
Semenoff, Phys. Lett. B {\bf 286}, 118 (1992).

\item{[16]}
More examples (in 1$d$ and 2$d$) and details will be published
elsewhere; see also cond-mat/0008374.

\item{[17]}
L. Huerta and J. Zanelli, Phys. Rev. Lett. {\bf 71}, 3622 (1993).

\item{[18]}
N. Papanicolaou, Nucl. Phys. B {\bf 305}, 367 (1988).

\item{[19]}
J.E. Gubernatis and N. Kawashima, in {\it Monte Carlo and Molecular
Dynamics of Condensed Matter Systems} (Societ\`a Italiana de Fisica,
Bologna, 1995) edited by K. Binder and G. Ciccotti.

\end